\documentclass[aps,pre,amsmath,amsfonts,floatfix,superscriptaddress]{revtex4}
\usepackage{axodraw2,color}
\usepackage{graphicx,graphics}
\usepackage{epstopdf}
\usepackage{amsmath}
\usepackage{rotating}
\usepackage{amssymb}
\usepackage{feynmp}
\usepackage{mathrsfs}
\usepackage{booktabs}
\input{epsf.sty}
\usepackage{bbm}

\let\a=\alpha \let\b=\beta  \let\g=\gamma  \let\d=\delta \let\e=\varepsilon
    \let\k=\kappa \let\l=\lambda
\let\m=\mu    \let\n=\nu         \let\p=\pi   
\let\s=\sigma \let\t=\tau   \let\f=\varphi 
   
\let\G=\Gamma \let\D=\Delta  \let\L=\Lambda 
       
 \let\ee=\epsilon \let\r=\rho \let\th=\theta
\let\io=\infty

\def\ba{\begin{align}}
\def\ea{\end{align}}

\def\MM{{\cal M}} 
 \def\WW{{\cal W}}
 \def\BB{{\cal B}}
\def\LL{{\cal L}}  
\def\DD{{\cal D}}\def\AA{{\cal A}}\def\GG{{\cal G}} 
\def\KK{{\cal K}}

\def\Tr{{\text{Tr}}}

    \def\hr{{\hat{\rho}}} \def\hrd{\hat{\rho}^{(2)}}
\def\rd{\rho^{(2)}} 

\def\to{\rightarrow}
\def\la{\left\langle}
\def\ra{\right\rangle}

\def\de{\mathrm d}

\newcommand{\beq}{\begin{equation}} \newcommand{\eeq}{\end{equation}}
 \newcommand{\wt}{\widetilde}

\begin{document}

\title{Static replica approach to critical correlations in glassy systems}

\author{Silvio Franz}
\affiliation{Laboratoire de Physique Th\'eorique et Mod\`eles
    Statistiques, CNRS et Universit\'e Paris-Sud 11,
UMR8626, B\^at. 100, 91405 Orsay Cedex, France}
\author{Hugo Jacquin}
\affiliation{Laboratoire Mati\`ere et Syst\`emes Complexes, UMR 7057,
CNRS and Universit\'e Paris Diderot -- Paris 7, 10 rue Alice Domon et L\'eonie 
Duquet, 75205 Paris cedex 13, France}
\author{Giorgio Parisi}
\affiliation{Dipartimento di Fisica,
Sapienza Universit\`a di Roma,
P.le A. Moro 2, I-00185 Roma, Italy}
\affiliation{
INFN, Sezione di Roma I, IPFC -- CNR,
P.le A. Moro 2, I-00185 Roma, Italy}
\author{Pierfrancesco Urbani}
\affiliation{Laboratoire de Physique Th\'eorique et Mod\`eles
    Statistiques, CNRS et Universit\'e Paris-Sud 11,
UMR8626, B\^at. 100, 91405 Orsay Cedex, France}
\affiliation{Dipartimento di Fisica,
Sapienza Universit\`a di Roma,
P.le A. Moro 2, I-00185 Roma, Italy}
\author{Francesco Zamponi}
\affiliation{LPT, Ecole Normale Sup\'erieure, UMR 8549 CNRS, 24 Rue Lhomond, 75005 France}

\begin{abstract} 
We discuss the slow relaxation phenomenon in glassy systems by means of replicas by constructing a static field theory approach to the problem. 
At the mean field level we study how criticality in the four point correlation functions arises because of the presence of soft modes and we derive 
an effective replica field theory for these critical fluctuations. By using this at the Gaussian level we obtain many physical quantities: the correlation 
length, the exponent parameter that controls the Mode-Coupling dynamical exponents for the two-point correlation functions, and the prefactor of the 
critical part of the four point correlation functions. Moreover we perform a one-loop computation in order to identify the region in which the mean field 
Gaussian approximation is valid. The result is a Ginzburg criterion for the glass transition. We define and compute in this way a proper Ginzburg number. 
Finally, we present numerical values of all these quantities obtained from the Hypernetted Chain approximation for the replicated liquid theory.
\end{abstract}

\maketitle

\tableofcontents

\section{Introduction}
Much progress in the recent understanding of glassy relaxation of
supercooled liquids has come from the study of dynamical
heterogeneities \cite{BBBCS11}. Space-time fluctuations of the density
field result in a distribution of regions of different mobility with
typical size that grows upon decreasing the temperature, and
that persist for time scales of the order of the relaxation
time. Remarkably, the present theories of glassy relaxation are able
to predict the qualitative features of this effect. Many progresses have been
achieved neglecting activated processes. Both theories based on
dynamics (Mode Coupling Theory~\cite{Go09}) and on constrained equilibrium
(e.g. the molecular liquid theory~\cite{MP09}) agree in a critical instability of the liquid
phase at a finite transition temperature. As far as universal aspects
are concerned, a first theoretical insight originally came from spin
glass theory, that suggested to look at the critical behavior of four
point density correlation functions \cite{KT88,FP00}. In the
context of p-spin models where both the equilibrium and the dynamic
approach can be pursued exactly, it was possible to predict the
qualitative features of dynamical correlations growth 
and associated dynamic criticality
observed in
numerical simulations~\cite{Gl97} 
to well defined features of static correlations in constrained
equilibrium. Beyond these schematic models a direct connection between
equilibrium and dynamics is more difficult to make. However the static
hallmarks of dynamical criticality are generically present whenever
one has a one step replica symmetry breaking transition (or mean-field
random first order transition~\cite{KT87b,WL12}), as it is found e.g. for liquids in the
Hypernetted Chain (HNC) approximation~\cite{MP96}.  At the level of dynamical liquid theory, the
growth of correlations has been found with generality within Mode
Coupling Theory \cite{BBMR06} through a one-loop diagrammatic expansion.
Despite these important results, in order to progress further it is necessary to build a field
theoretical description of dynamical fluctuations capable in principle
to go systematically beyond the zero-loop Gaussian level. Unfortunately, the dynamic
approach is rather problematic in this respect for several reasons. The
MCT is not a self-consistent theory as it requires the input of the
static structure factor. Moreover, the equation for the dynamic
correlator does not follow from a variational principle. Clearly this
is not the ideal starting point to build up a perturbative computation. In
addition an expansion around mean field in dynamics is prohibitively
difficult even in the simplest cases.

A static formulation would be therefore very useful to make additional progresses.
The connection between statics and dynamics is based on the idea
that the emergence of slow dynamics is due to the appearance of long
lived metastable states~\cite{KW87,KT87}. Within Mode-Coupling Theory
and similar dynamical theories of glass formation, one can distinguish two separate
dynamical regimes: the so-called $\b$ regime, that corresponds to the long time dynamics
inside a metastable state, and the $\a$ regime, that corresponds to transitions between 
different metastable states.
A general theory of fluctuations in the $\beta$ regime based on the
replica method was proposed in \cite{FPRR11} on the basis of general
symmetry considerations.
It was found that the dynamical
transition is in the universality class of a cubic random field Ising model. 
Moreover in \cite{CFLPRR12} it has been shown at the level of schematic models that the same theory can be used to evaluate the 
Mode-Coupling dynamical exponents and that  they are related to the amplitude ratio between correlation
functions that are in principle measurable.
At the Gaussian level, we can consider this phenomenological theory as the static Landau theory of the glass transition
(see~\cite{ABB09} for a discussion of a dynamical Landau theory of the glass transition).
In summary, the advantage of the approach discussed in~\cite{FPRR11} is that one can compute the long time behavior 
of the dynamical correlations in the $\beta$ regime starting from a completely static replicated (and constrained) Boltzmann measure~\cite{KT89,Mo95, FP95}.
The first consequence of this is that all the calculations simplify drastically. 
Moreover one can look at the dynamical transition through a static measure and one can see systematically where the mean field prediction is valid and where it fails.

The aim of this paper is to construct a framework to study the glass transition in the $\beta$ regime by 
following as closely as possible the standard treatment of critical phenomena using field theoretical tools.
Furthermore, we want to obtain the replica field theory of critical fluctuations starting directly from the 
microscopic grand-canonical expression for the replicated partition function for the liquid, in such a way that the couplings
appearing in the field theory can be computed starting from the microscopic potential.
This program can be achieved through an analysis of the soft modes that appear at the dynamical transition 
and that are responsible for the criticality. By computing these soft modes we can focus on fluctuations that are 
along them:
in this way we construct a gradient expansion for the field theory of the critical fluctuations and we compute in full details the 
Gaussian correlation functions, thus obtaining the critical part of the long time limit of the dynamical four point functions in the $\b$ regime.
Then we look at the corrections to the Gaussian theory and we introduce a Ginzburg criterion 
in order to see where the Gaussian theory is valid. The Ginzburg criterion can be used in two ways. 
On the one hand it gives the upper critical dimension for the model, on the other hand it provides a measure of
how much one has to be close to the transition line in order to see the non Gaussian fluctuations that cannot be treated by the mean field approach.

A short account of our findings appeared in~\cite{FJPUZ12}.
This paper is organized as follows. In Section \ref{sec:ill} we review all the line of reasoning we just discussed for the standard Ising model. 
Then, in Section III we discuss how to obtain a replica description for the dynamical correlation functions in the $\beta$ regime. 
In this way we rephrase the problem from a dynamical one to a standard static computation.
In Section IV we discuss the expansion of the free energy around the glassy solution. By studying the Hessian matrix we identify the soft modes that appear at the dynamical transition and we compute the expression for the exponent parameter $\l$ that can be related to the dynamical Mode-Coupling exponents $a$ and $b$ that describe the critical slowing down of the two point correlation functions at the dynamical transition. Then, in Section V we show how we can perform a gradient expansion of the replica field theory in order to study the long distance physics and in Section VI we use this effective theory to compute where the Gaussian level computation fails by introducing a Ginzburg criterion for the dynamical transition.
Up to that point we will remain completely general. The only assumption that we will make is that the replicated system has a glassy phenomenology, namely that the replica structure of the two point density function is non trivial below a certain dynamical transition point. In Section VII we report concrete calculations in the framework of the replicated HNC approximation by giving explicit expressions for all the couplings and masses of the effective replica field theory starting from the microscopic potential and we show the numerical results for several physical systems.

\section{An illustration of the main results of this paper in the simpler case of a standard ferromagnetic transition}
\label{sec:ill}

The aim of this paper is to build a theory of the glass transition by following closely the first steps 
of the standard field theory formulation of critical phenomena. Namely, we want to start from a Landau
theory (including microscopic parameters) and deduce from it a set of mean field critical exponents.
This is done first by studying the behavior of the uniform order parameter in the mean field theory, and
then considering a gradient expansion for slowly varying order parameter to compute the correlation length.
Finally, taking into account the interaction terms lead to a perturbative loop expansion that allow to establish
the region of validity of mean field theory (hence the upper critical dimension) through a Ginzburg criterion. 
The aim of this paper is to repeat all
of these steps in the more complex case of a system undergoing a glass transition. The final step would be
of course to set up an epsilon expansion around the upper critical dimension using renormalization group
methods. We will not make any attempt in this direction in this paper.
For pedagogical reasons, we find useful to briefly describe how these steps are carried out in a simple
ferromagnetic system before turning to the glass case. The reader should keep in mind that
this section is just a short reminder
of the main steps, for a reader already accustomed with the modern theory of critical phenomena.

\subsection{Landau theory}
\label{sec:illA}

Suppose we consider a microscopic system undergoing a ferromagnetic transition.
In the following we will consider the ferromagnetic Ising model
with nearest neighbor interactions on a $D$-dimensional cubic lattice and a
properly scaled coupling constant:
\beq\label{eq:HIsing}
 H[\s] = -\frac{1}{2 D} \sum_{\la i,j \ra} \s_i \s_j \ .
\eeq
Starting from the microscopic Hamiltonian, we can construct the free
energy as a functional of the order parameter, the magnetization $\phi_i = \la \s_i \ra$, 
as follows.
The free energy (here we will ignore some factors of temperature by rescaling some variables) 
in presence of an external magnetic field is
\beq
W[h] = \log Z[h] = \log \sum_{\{\s_i = \pm 1\}} e^{-\b H[\s] + \sum_i h_i \s_i} \ .
\eeq
Taking a Legendre transform~\cite{CJT74}, we define
\beq
\G[\phi] = \sum_i h^*_i \phi_i - W[h^*] \ ,
\eeq
or in other words
\beq\label{eq:GammaLegendre}
e^{-\G[\phi]} = \sum_{\{\s_i = \pm 1\}} e^{-\b H[\s] + \sum_i h^*_i (\s_i - \phi_i)} \ ,
\eeq
where $h^*$ is the solution of $\frac{dW[h]}{dh_i} = \phi_i$. 
The function $\G[\phi]$ is the free energy of the system as a function of the magnetization field.

In order to detect the phase transition, we want to investigate the small $\phi$ behavior
of $\G[\phi]$. 
Let us make a crucial assumption,
{\it that $\G[\phi]$ is an analytic function of $\phi$ around $\phi=0$}.
This assumption is plain wrong in finite dimensional systems at the critical point and below.
However, let us for the moment forget about this problem and proceed with our discussion.
We can consider a uniform magnetization profile $\phi$ and expand $\G[\phi]$ at small $\phi$.
From the symmetries of the problem, we know that
\beq\label{Landau_ferro}
\G[\phi] = V \left\{ \frac12 m_0^2 \phi^2 + \frac{g}{4!} \phi^4 + \cdots \right\} \ ,
\eeq
which is the celebrated Landau free energy (here $V$ is the volume of the system).
A very practical way to compute the coefficients $m_0^2$ and $g$ is to perform a systematic
high temperature expansion~\cite{GY91}.
For example, at the leading order, for the $D$-dimensional Ising model with coupling constant
$J=1/(2D)$ we obtain $m_0^2 = 1 - \b$. Adding more terms leads to an expansion of $m_0^2$ in
powers of $\b$. 
For the Ising model \eqref{eq:HIsing}, the true expansion parameter is actually $\b J$, i.e.
the temperature in units of the coupling constant. Because the latter has to be chosen equal
to $J=1/(2D)$ to obtain a good limit $D\to\io$, the expansion parameter is $\b/(2D)$. In other
words, the high temperature expansion is also a large dimension expansion around the $D=\io$
mean field limit.

The equilibrium value of $\phi$, called $\overline{\phi}$, is obtained by minimizing $\G[\phi]$.
From the high temperature expansion we obtain that $m_0^2$ vanishes linearly 
at a given temperature
$T_c$,
in such a way that $m_0^2 \propto T/T_c-1 = \ee$.
Note that for instance in $D=3$ it is enough to consider the cubic term in the small $\b$ expansion
to obtain a fairly accurate estimate of $T_c$.
When $m_0^2$ becomes negative, the magnetization becomes non-zero
with $\overline{\phi} \sim | m_0^2 |^{1/2} \sim \ee^{1/2}$ 
which gives one of the critical exponents. The other critical 
exponents are obtained
in a similar way.

\subsection{Gradient expansion}
\label{sec:illB}

The subsequent step is to compute the correlation length. This is done by considering a gradient expansion
for a slowly varying magnetization profile, again under the 
assumption that the expansion is regular at small $\phi$.
One can perform a continuum limit to simplify the notations: 
we denote by $\f(x)$ the continuum limit of the spin field $\s_i$, while
$\phi(x) = \la \f(x) \ra$ is the local average magnetization.
The Landau free energy becomes
at the quadratic order:
\beq\label{Landau_ferro_gradient}
\G[\phi]=\frac{1}{2}\int \de x\,\phi(x)(-\nabla^2+m_0^2)\phi(x) \ .
\eeq
The correlation function of the magnetization is given by 
\beq
G(x-y) = \la \f(x) \f(y) \ra = 
\left[ \frac{\partial^2 \G}{\partial \phi(x) \partial \phi(y) } \right]^{-1} \ .
\eeq
Hence at this order the correlation function
is
\beq\label{G0ferro}
G_0(p) =  \frac{1}{p^2 + m_0^2} \ , \hskip30pt
G_0(x) = \la \f(x) \f(0) \ra \sim_{x\to\io} x^{\frac{4-D-3}2} e^{-m_0 x} \ ,
\eeq
which is often called the {\it bare propagator}.
The calculation is performed by using $1/(p^2+m_0^2) = \int_0^\io dt \, e^{-(p^2 + m_0^2) t }$ and
changing variable to $y=t/x^2$. Then
\beq
G_0(x) \propto x^{2-D} \int_0^\io dy \, e^{-(m_0 x)^2 y - \frac1{4y}} y^{-D/2} = x^{2-D} f(m_0 x) \ .
\eeq
When $x \gg 1/m_0$, a saddle point calculation shows that $f(z) \sim z^{(D-3)/2} e^{-z}$, from
which Eq.~\eqref{G0ferro} follows.
This expression shows that the correlation length is 
$\xi  =1/m_0 \sim \ee^{-1/2}$ and the magnetic susceptibility
is $\chi \propto G_0(p=0) \sim \ee^{-1}$.

\subsection{From the microscopic Hamiltonian to a field theoretical formulation}
\label{sec:illC}

The above analysis relies on the assumption that $\G[\phi]$ can be expanded as an analytic
function around $\phi=0$. 
Although this is certainly true at the mean field level (where the Landau theory provides
the exact result), this is not the case in finite dimensional systems, because
critical fluctuations induce a singular behavior of $\G[\phi]$ at small $\phi$.
Hence we now want to assess the limits of validity of the Landau theory by studying the effect
of critical fluctuations.

The problem is that the definition of $\G$ given in Eq.~\eqref{eq:GammaLegendre} is
not very convenient to perform a systematic expansion in the fluctuations around the mean
field theory, although the computation could be done in principle.
It would be much more convenient to write the effective action as a functional integral over
a continuous spin field $\f(x)$:
\beq\label{eq:GammaLegendreS}
e^{-\G[\phi]} = \int \DD\f \, e^{-S[\f] + \int \de x h(x) [\f(x) - \phi(x)]} \ ,
\eeq
with the following requirements:
\begin{enumerate}
\item The mean field approximation should correspond to a saddle point
evaluation of the above integral, in such a way that at the mean field level
$\G[\phi] = S[\phi]$. For consistency, $S[\f]$ must therefore have a Landau form:
\beq\label{eq:Sbare1}
S[\f]=\frac{1}{2}\int \de x\,\f(x)(-\nabla^2+m_0^2)\f(x)+\frac{g}{4!}\int\de x \f^4(x) \ ,
\eeq
in such a way that at the mean field level we 
recover Eqs.~\eqref{Landau_ferro} and \eqref{Landau_ferro_gradient}. 
In this way we can include
fluctuations around mean field by performing a systematic loop expansion of the functional
integral.
\item The bare coefficients $m_0^2$ and $g$ entering in $S[\f]$ must be reasonable approximations
to the microscopic coefficients as deduced from the Hamiltonian, in such a way that the mean
field approximation is already a good approximation, and that loop corrections improve systematically
over it. In this way we can guarantee that the criterion of validity of mean field theory has
a quantiative meaning for the original microscopic Hamiltonian $H[\s]$.
\end{enumerate}
So we want to give an appropriate definition of the continuum spin field $\f(x)$ and the corresponding
action $S[\f(x)]$ in such a way that the requirements above are satisfied.

There are many recipes for such a construction. Probably the best one is given by the
non-perturbative renormalization group construction~\cite{De07}, 
in which one defines a functional
$\G_\ell[\f(x)]$ by integrating the small-scale spin fluctuations on length scales smaller than
$\ell$, see e.g. \cite[Eq.~(28)]{De07}. 
If one chooses a ``coarse-graining'' length $\ell_0$ 
that is quite bigger than the lattice spacing, but still quite small
with respect to the correlation length (which diverges at the critical point), the function
$\G_\ell[\f(x)]$ is an analytic function of $\f$ at small $\f$, because the singularity is only
developed at the critical point for $\ell\to\io$~\cite{De07}. 
Then, Eq.~\eqref{eq:GammaLegendreS} is basically exact with $S$ replaced by $\G_{\ell_0}$.
One can then expand $\G_{\ell_0}$ at small
$\f$ and use this as the bare action $S$ in Eq.~\eqref{eq:GammaLegendreS}.
Although this strategy can be generalized to the physics of liquids~\cite{PR85,Ca06}, 
calculations are quite involved so we need to consider something simpler.

An alternative and very convenient prescription is the following. Let us call
$\G_k$ the truncation at a finite order $\b^k$ of the high temperature expansion
of $\G$, as given in~\cite{GY91}.
We know that
$\G_k[\phi]$ is an analytic function of $\phi$ for any finite $k$, hence
$\G_k[\phi]$ cannot be a good approximation of $\G[\phi]$
at the critical point, because we know that $\G[\phi]$ is not analytic: 
in fact the high temperature expansion is divergent at the critical point.
However, we can assume that 
{\it $\G_k[\f]$ gives a good approximation for $S[\f]$.}
Note that our two requirements are satisfied by the prescription that
$S[\f] = \G_k[\f]$. In fact, for the first requirement,
at the saddle point level we obtain $\G[\phi] = S[\phi] = \G_k[\phi]$, and we already
know that for $k=1$ this is the correct mean field result, while for $k>1$ we will
obtain an ``improved'' mean field result.
For the second requirement, we have already mentioned that the coefficients
of $\G_k[\phi]$ give, for large enough $k$, a good estimate of the microscopic properties
of the model (e.g. the critical temperature).
Furthermore, we can argue that the high temperature expansion, at a given order $k$, is only
sensitive to local physics up to a scale $\ell(k)$ that grows with $k$. Hence, truncating
the high temperature expansion at a finite order in $k$ should be morally equivalent
to perform an integration over the microscopic fluctuations on a scale smaller than
$\ell(k)$. We will see that this procedure is easily generalized to the case of liquids
where the high temperature expansion is replaced by the low-density virial expansion.

We will then use the prescription $S[\f] = \G_k[\f]$, expand $S[\f]$ in the
form of Eq.~\eqref{eq:Sbare1}, and use it in the functional integral 
Eq.~\eqref{eq:GammaLegendreS} to compute $\G[\phi]$ in a loop expansion around mean field.
Loop corrections give some non-singular contributions to $\G$, which were already in part taken
into account in the bare action $S[\f] = \G_k[\f]$ because it was obtained from the high temperature
expansion: hence we might have some ``double counting'' of non-singular contributions related to the
short range physics. This double counting problem is discussed in more details in Appendix~\ref{app:DC}.
Still, our aim here is to find a Ginzburg criterion that identifies
the region where these singular loop corrections are small, and the mean field approximations remains
correct: we find that if the Ginzburg criterion is formulated in terms of physical quantities, then
double countings are irrelevant. This is shown in next section~\ref{sec:illD} and in Appendix~\ref{app:DC}.

\subsection{Ginzburg criterion}
\label{sec:illD}

We can use the above construction to perform a loop expansion in the coupling and
check whether fluctuations are small such that they
do not spoil the main assumptions we made above on the behavior of $\G$ at small $\phi$,
hence they do not change the critical behavior of the system.
We will follow closely the derivation of~\cite{Am74}.
Our bare action is
\beq
S[\f]=\frac{1}{2}\int \de x\,\f(x)(-\nabla^2+m_0^2)\f(x)+\frac{g}{4!}\int\de x\f^4(x) \ .
\eeq
Here we will need to consider explicitly the presence of an ultraviolet cutoff (which will be of the
order of the scale $\ell_0(k)$ mentioned above).
The bare propagator is
\beq\label{barepropcount}
G_0(p) =  \frac{1}{p^2 + m_0^2} \ ,
\eeq
and the one loop correction to the propagator is~\cite{ParisiBook}
\beq\label{1Lprop}
G(p) = G_0(p) -  \frac{g}{2} G_0(p)^2  \int^\L \frac{\de q}{(2 \pi)^D} G_0(q) \ .
\eeq
We can consider the inverse propagator and write
\beq
G^{-1}(p) = G_0(p)^{-1} +  \frac{g}{2}  \int^\L \frac{\de q}{(2 \pi)^D} G_0(q) \ ,
\eeq
which can be seen either as a Dyson resummation of the ``tadpole'' diagrams, 
or as an inversion of the perturbation expansion to obtain 
directly the second derivative of the Legendre transform
of the generating functional. Physically, $G^{-1}(p=0)$ is the ``renormalized mass'' or inverse magnetic susceptibility:
\beq\label{mRm0}
m^2_R = G^{-1}(p=0) = m_0^2 + \frac{g}2  \int^\L \frac{\de q}{(2 \pi)^D}  \frac{1}{q^2 + m_0^2 } = m_0^2 
+ \frac{g}2  \int^\L \frac{\de q}{(2 \pi)^D}  \frac{1}{q^2 + m_R^2 }  \ ,
\eeq
where the last equality of course holds at first order in $g$. The replacement of $m_0$ by $m_R$ is needed,
because the perturbation theory must be done at fixed $m_R$, i.e. at fixed distance from the true critical point,
otherwise corrections cannot be small~\cite{ParisiBook}.
The critical point is defined by the condition that $m_R^2=0$, or in other words the susceptibility
is divergent, hence at the critical point
\beq
m_0^2 =  - \frac{g}{2}  \int^\L \frac{\de q}{(2 \pi)^D}  \frac{1}{q^2 } \ .
\eeq
We see that the shift of the critical temperature is divergent in the ultraviolet (UV divergent) for $D\geq 2$:
indeed, this is a non-universal quantity and
depends on the details of the UV regularization. 
Now if we define the distance from the critical point as
\beq
t = m_0^2 + \frac{g}{2}  \int^\L \frac{\de q}{(2 \pi)^D}  \frac{1}{q^2 } \ ,
\eeq
we can write Eq.~\eqref{mRm0} as
\beq
t = m_R^2 \left( 1   - \frac{g}{2}  \int^\L \frac{\de q}{(2 \pi)^D}  \frac{1}{q^2 (q^2 + m_R^2) } \right) \ .
\eeq
This is the crucial relation that relates the inverse susceptibility to the distance from the critical point.
The Ginzburg criterion is obtained by imposing that the one loop corrections do not change the mean field behavior
$t = m_R^2$.

We can now distinguish two cases:
\begin{itemize}
\item
For $D < 4$, the correction is UV convergent and infrared (IR) divergent. 
In this case, we can safely send the cutoff to infinity because the renormalized theory exists.
We obtain
\beq
t = m^2_R \left( 1 - \frac{g}{2}  \int^\io \frac{\de^D q}{(2 \pi)^D}  \frac{1}{q^2 (q^2 + m_R^2) } \right) = 
m^2_R - \frac{g}{2} \, m_R^{ D-2 } \frac{\Omega_D}{(2 \pi)^D}  \int_0^\io \de x \, x^{D-1} \frac{1}{x^2 (x^2 + 1) } 
\eeq
and the integral over $x$ is finite. 
We clearly see that because $D<4$, the second term will be dominant over the first close enough to the critical point.
By imposing that the first term dominates, we obtain the criterion in the form
\beq
1 \gg g \, m_R^{ D-4 } C_D = g \xi^{4-D} C_D = \text{Gi} \, \xi^{4-D} \ ,
\eeq
where we used that in the mean field region the correlation length $\xi = 1/m_R$.
Hence the Ginzburg number $\text{Gi} = g C_D$ is a universal constant in this case.
This shows that loop corrections will always be relevant close enough to the critical point
and gives a precise value of the correlation length at which they will become relevant,
$\xi \sim 1/(\text{Gi})^{1/(4-D)}$.
\item
Instead, for $D \geq 4$, the correction is UV divergent and IR convergent. 
In this case the Ginzburg criterion is non-universal and strongly dependent on the 
details of the regularization. For a fixed cutoff $\L$, the integral is finite at $m_R^2=0$ and
the mean field behavior is always correct:
\beq
t = m_R^2 \left( 1   - \frac{g}{2}  \int^\L \frac{\de q}{(2 \pi)^D}  \frac{1}{q^4 } \right) \ .
\eeq
However one loop corrections provide a strong renormalization of the coefficient relating $t$
to $m_R^2$. Imposing that these conditions are small we obtain
\beq
1 \gg  \frac{g}{2}  \int^\L \frac{\de^D q}{(2 \pi)^D}  \frac{1}{q^2 (q^2 + m_R^2) }
\eeq
This provides a condition on $m_R$ for a given UV cutoff $\L$. When the condition is satisfied
the mean field calculation is not only qualitatively, but also quantitatively correct.
Note that the integral is upper-bounded by its value in $m_R=0$. 
Then, if $1 \gg g C \L^{D-4}$, the Ginzburg criterion is always satisfied and one loop
corrections are small even at the critical point. Instead, if $g C \L^{D-4}$ is of order 1 of bigger, 
then we obtain a non-trivial condition on $m_R$ and one loop corrections are large close enough
to the critical point.
\end{itemize}

\section{Dynamical heterogeneities and replicas: definitions}
\label{sec:rep}

The aim of this paper is to repeat the steps outlined in Sec.~\ref{sec:ill} in the case of a glass transition. As we will see, the calculation is in this case complicated by
several problems:
\begin{enumerate}
\item We will need to introduce replicas to define a proper static order parameter.
\item The order parameter is in general not a real number (e.g. the magnetization) but a {\it function} $\wt g(x-y)$ that encodes the replica-replica correlations.
Hence we will have to introduce a smoothing function to define a scalar order parameter $q \sim \int f(x) \wt g(x)$. We will then have to show that the choice of
the function $f(x)$ is irrelevant.
\item The glass transition is discontinuous in the order parameter, which jumps to a finite 
value at the transition. Hence the transition is not an instability of the high temperature
solution, but rather a {\it spinodal} point where the low temperature solution first appear. 
Because of that, we need to control the effective free energy at values of the order parameter
that are very far away from the high temperature solution. Keeping only a few terms in the
high temperature expansion is not enough, and we will have to resum an infinite number of
terms to obtain a good starting point for the mean field theory.
\item 
Because the glass transition is akin to a spinodal point, 
the resulting effective action is a cubic theory. 
Hence the theory is not really defined
(spinodals do not exist in finite dimension). This will not be a problem for the mean field and loop calculations, but we expect it to be a serious problem
if one wants to go beyond mean field and construct a systematic epsilon expansion (which we do not attempt here).
\end{enumerate}
In this section we will better explain the first two points: we will
give some important basic definitions on criticality at the glass transition (as encoded by the so-called {\it dynamical heterogeneities}) 
and we will show how the problem can be tackled using replicas.
In Sec.~\ref{sec:defA} we introduce the basic dynamical order parameter of the glass transition, and in Sec.~\ref{sec:defB} its correlation function.
In Sec.~\ref{sec:defC} we show how both quantities can be written as static correlations in a replicated theory.
In Sec.~\ref{sec:defD} we set up the general form of this replicated theory and give some useful definitions.

Throughout this paper we consider a system of $N$ particles in a volume $V$ interacting through a pairwise potential $v(r)$ in a $D$ dimensional space.
The basic field is the local density at point $x$ and time $t$:
\beq
\hat\r(x,t) = \sum_{i=1}^N \d(x-x_i(t)) \ .
\eeq
We will consider a generic dynamics that can be either Newtonian or stochastic, e.g. of Langevin type.
In both cases, we will consider {\it equilibrium} dynamics, that starts from an equilibrated configuration of the system.
It will be convenient to separate the dynamical average in two contributions~\cite{FPRR11}. A dynamical history of the system will be 
specified by the initial configuration of the particles $\{x_i(0)\}$, and by a dynamical noise. For Newtonian dynamics, this noise
comes from the initial values of the velocities, extracted by a Maxwell distribution; for stochastic dynamics, it comes from the 
random forces that appear in the dynamical equations.
We will denote by $\la \bullet \ra$ the average over the dynamical noise for a fixed initial condition; and by $\mathbf E[ \bullet ]$ the
average over the initial condition. Hence, for instance, the equilibrium average of the density will be
$\r = \mathbf E[ \la \hat\r(x,t) \ra]$.

\subsection{Two-point functions: the dynamical order parameter}
\label{sec:defA}

The dynamical glass transition is characterized by an (apparent) divergence of the relaxation time of density fluctuations,
that become frozen in the glass phase.
The transition can be conveniently characterized using correlation
functions. Consider the density profiles 
at time zero and 
at time $t$, respectively given by 
$\hat\rho(x,0)$ and $\hat\rho(x,t)$. We can define a local similarity measure of these 
configurations as 
\beq\label{ove}
\begin{split}
\hat C(r,t) &= \int \de x f(x) \hat\r\left(r+\frac{x}2,t\right) \hat\r\left(r-\frac{x}2,0\right) \\
&=\sum_{ij} \d\left( r- \frac{x_i(t)+x_j(0)}2 \right) \, f(x_i(t) - x_j(0))  \ .
\end{split}\eeq
Here $f(x)$ is an arbitrary ``smoothing"
function of the density field with some short range $a$, which is normalized in such a way that $\int \de x f(x) =1$.
As an illustration, let us choose $f(x) = \th(a-|x|)/V_d(a)$, where $V_d(a)$ is the volume of a sphere of radius $a$, and
suppose that $a$ is much smaller than the inter-particle distance and that $t$ is short enough.
In this situation, $f(x_i(t) - x_j(0))$ vanishes unless $i=j$, and we get
\beq\label{Cself}
\begin{split}
\hat C(r,t) \approx \sum_{i} \d\left( r- \frac{x_i(t)+x_i(0)}2 \right) \, f(x_i(t) - x_i(0))  \ .
\end{split}\eeq
Therefore, $\hat C(r,t)$ counts how many particles that are around point $r$ have moved less then $a$ in time $t$ and is often
called ``mobility'' field. 
Alternatively, Eq.~(\ref{Cself}) can be taken as the definition of a {\it self} two-point correlation function.
Different choices of $f(x)$ lead to other correlations that have been used in different studies. We will show
later that the choice of the function $f(x)$ is irrelevant as far as the critical properties are concerned.

Let us call 
\beq
C(t) =V^{-1} \int \de r \mathbf E[\langle \hat C(r,t) \rangle] - \r^2
\eeq
the spatially and thermally averaged connected correlation function.
Typically, on approaching the dynamical glass transition $T_{\rm d}$, 
$C(t)$ displays a two-steps relaxation, with a fast ``$\b$-relaxation'' 
occurring on shorter times down to a ``plateau'',
and a much slower ``$\a$-relaxation'' from the plateau to zero~\cite{Go09}.
Close to the plateau at $C(t) = C_{\rm d}$,
one has $C(t) \sim C_{\rm d} + \AA \, t^{-a}$ in the $\b$-regime.
The departure from the plateau (beginning of $\a$-relaxation)
is described by $C(t) \sim C_{\rm d} - \BB \, t^b$.
One can define the $\a$-relaxation time by $C(\t_\a) = C(0)/e$. 
It displays an apparent power-law divergence at the transition, 
$\t_\a \sim |T-T_{\rm d}|^{-\g}$.
All these behaviors are predicted by MCT~\cite{Go09}, which in particular relates
all these exponents to a single parameter $\l$ through the formulae
\beq\begin{split}
&\frac{\G(1-a)^2}{\G(1-2a)} = \frac{\G(1+b)^2}{\G(1+2b)} = \l \ , \\
&\g = \frac1{2a} + \frac1{2b} \ ,
\end{split}\eeq
and gives a microscopic expression of $\l$ in terms of liquid correlation functions~\cite{Go09}.
In low dimensions, a rapid crossover to a different regime dominated by activation 
is observed and the divergence at $T_{\rm d}$ is
avoided; however, the power-law regime is the more robust the higher the dimension~\cite{CIMM10,CIPZ12}
or the longer the range of the interaction~\cite{IM11}.

\subsection{Four point functions: the correlations of the order parameter}
\label{sec:defB}

It is now well established, both theoretically and experimentally, 
that the dynamical slowing is accompanied by growing heterogeneity of the local relaxation, in the sense that the local
correlations $\hat C(r,t)$ display increasingly correlated fluctuations when $T_{\rm d}$ is approached~\cite{FP00,DFGP02,BBBCEHLP05,BBBCS11}.
This can be quantified by introducing the correlation function of $\hat C(r,t)$, i.e. a four-point dynamical correlation
\beq\label{G4}
\begin{split}
&G_4(r,t)=
\mathbf E[  \langle  \hat C(r,t) \hat C(0,t) 
\rangle] - \mathbf E[\langle 
\hat C(r,t)\rangle] \mathbf E[\langle \hat C(0,t) 
\rangle ] \ .
\end{split}\eeq
This function describes the total fluctuations of the two-point correlations, and it 
decays as $G_4(r,t)\sim \exp(-r / \xi(t))$ with a ``dynamical correlation length'' 
that grows at the end of the $\b$-regime and
has a maximum $\xi = \xi(t\sim \t_\a)$
that also (apparently) diverges as a power-law when $T_{\rm d}$ is approached.
MCT~\cite{Go09} and its extensions~\cite{BBMR06,BB07,BBBKMR07a,BBBKMR07b,Sz08,SF10} 
give precise predictions for the critical exponents.

For later convenience, we can also consider a modified four-point correlation:
\begin{gather}
G_{th}(r,t)=\mathbf E\left[  \langle  \hat C(r,t) \hat C(0,t) 
\rangle - \langle 
\hat C(r,t)\rangle  \langle \hat C(0,t)  \rangle   \right] \ .
\end{gather}
This function describes the {\it isoconfigurational} fluctuations of the two-point correlations, i.e. the fluctuations due
to the noise of the dynamical process at fixed initial condition.
It describes the in-state susceptibility: indeed, the initial condition selects a typical 
glass state, which is then explored by the dynamics.
A final average over initial conditions is taken to ensure that the initial condition is a typical one.

For each of these correlations, we can define the corresponding susceptibility
\beq\begin{split}
\chi_4(t) &= \int \de r G_4(r,t) = \mathbf E\left[  \left\langle  \left( \frac{1}{V} \int \de r \hat C(r,t) \right)^2
\right\rangle \right] - \mathbf E\left[ \left\langle 
\frac{1}{V} \int \de r \hat C(r,t) \right\rangle \right]^2 \ ,  \\
\chi_{th}(t) &= \int \de r G_{th}(r,t) = \mathbf E \left[  \left\langle  \left( \frac{1}{V} \int \de r \hat C(r,t) \right)^2 \right\rangle
 - \left\langle 
\frac{1}{V} \int \de r \hat C(r,t) \right\rangle^2 \right] \ .
\end{split}\eeq

 \subsection{Connection between replicas and dynamics}
 \label{sec:defC}

The dynamical glass transition can be also described, at the mean field level, in a static framework.
This has the advantage that calculations are simplified so that the theory can be pushed much forward, in particular by constructing a reduced
field theory and setting up a systematic loop expansion that allows to obtain detailed predictions for the upper critical dimension and the critical
exponents~\cite{FPRR11}. Moreover, very accurate approximations for the static free energy of liquids have been constructed~\cite{Hansen}, and one
can make use of them to obtain quantitative predictions for the physical observables. 

 In the mean field scenario, the dynamical transition of MCT is related to the emergence of a large number of metastable states
 in which the system remains trapped for an infinite time. At long times in the glass phase, the system is able to decorrelate within
 one metastable state. The $\beta$ regime is identified with the dynamics ``inside a metastable state'', while the $\a$ regime
 is identified with ``transitions between different states''.
 Hence we can write (introducing two new averages):
 \beq\label{Cmeta}
 \begin{split}
 \langle  \hat C(r,t \to\io) \rangle &= \int \de x f(x) \la \hat\r\left(r+\frac{x}2 \right) \ra_{\rm m} \la \hat\r\left(r-\frac{x}2\right) \ra_{\rm m} \ , \\
\mathbf E[\langle  \hat C(r,t \to\io) \rangle] &= \int \de x f(x)  \overline{\la \hat\r\left(r+\frac{x}2 \right) \ra_{\rm m} \la \hat\r\left(r-\frac{x}2\right) \ra_{\rm m}} \ .
\end{split} 
\eeq
In fact, if one performs a dynamical average at fixed initial condition, the system is trapped in a single metastable that can be explored, and at long
times the dynamical average can be replaced by the average $\langle \bullet \rangle_{\rm m}$ in the metastable state selected by the initial condition.
The average over the initial condition then induces an average over the metastable
states with equilibrium weights, that we denoted by an overline.

For the four-point functions we obtain
\beq\label{Gio}
\begin{split}
G_4(r,t\to\io) &= \int \de x \de y \, f(x) f(y) \, [ \overline{\langle \hat\r(r-x/2) \hat\r(-y/2) \rangle_{\rm m} \langle \hat \r(r+x/2)
  \hat \r(y/2) \rangle_{\rm m} } ] \\
&- \mathbf E[\langle 
\hat C(r,t\to\io)\rangle] \mathbf E[\langle \hat C(0,t\to\io) 
\rangle ] \ , \\
G_{th}(r,t\to\io) &= \int \de x \de y \, f(x) f(y) \, [ \overline{\langle \hat\r(r-x/2) \hat\r(-y/2) \rangle_{\rm m} \langle \hat \r(r+x/2)
  \hat \r(y/2) \rangle_{\rm m} } ] \\
&- 
\int \de x \de y \, f(x) f(y) \, [ \overline{\langle \hat\r(r-x/2) \rangle_{\rm m} \langle \hat\r(-y/2) \rangle_{\rm m} \langle \hat \r(r+x/2)
  \hat \r(y/2) \rangle_{\rm m} } ] 
 \ . \\
\end{split}\eeq
The reason for the particular structure of the second term of $G_{th}$ is that the densities at time 0 come from the same initial condition
and are therefore correlated, but they then evolve separately and therefore the two densities at time $t$ are uncorrelated.

The above structure suggests that
the dynamical transition can be described in a static framework
by introducing a replicated version of the system~\cite{KT89,Mo95,MP96}.
In fact, the replica method allows exactly to compute averages of the form $\overline{\la \bullet \ra_{\rm m}\la \bullet \ra_{\rm m}}$,
that enter in Eq.~(\ref{Cmeta}), in a static framework without the need of solving the dynamics.
For every particle we introduce $m-1$ additional particles identical to the first one. 
In this way we obtain $m$ copies of the original system, labeled by $a=1,\ldots,m$.
The interaction potential between two particles belonging to replicas $a,b$ is $v_{ab}(r)$.
We set $v_{aa}(r) = v(r)$,
the original potential, and we fix $v_{ab}(r)$ for $a\neq b$ to be an attractive potential that constrains
the replicas to be in the same metastable state.

Let us now define our basic fields that describe the one and two point density functions
\beq\label{physical_fields}
\begin{split}
&\hat\rho_a(x)=\sum_{i=1}^N\delta(x-x_i^{a}) \ , \\
&\hat\rho_{ab}^{(2)}(x,y)=\hat\rho_a(x)\hat\rho_b(y)-\hat \rho_a(x)\delta_{ab}\delta(x-y) \ .
\end{split}
\eeq
To detect the dynamical transition one has to study the two point correlation functions when
$v_{ab}(r)\to 0$ for $a\neq b$, and in the limit $m\to 1$ 
which reproduces the original model~\cite{Mo95,MP96}.
We denote by $\la \bullet \ra_{\rm r}$ the equilibrium average for the replicated system under
the conditions stated above.
The crucial observation is that in the limit $v_{ab}(r)\to 0$, all replicas fall in the same state 
but are otherwise uncorrelated inside the state.
This leads to the following rule to compute the average $\la \bullet \ra_{\rm r}$: one should
\begin{itemize}
\item
replace $\la \bullet \ra_{\rm r} = \overline{\la \bullet \ra_{\rm m}}$,
\item 
factorize the averages $\la \bullet \ra_{\rm m}$ when they involve different replicas, and
\item
remove the replica indexes.
\end{itemize}
For instance, for any spatial argument, and for $a\neq b$, we have that following the prescription above
\beq
\la \hat\r_a \hat\r_b \ra_{\rm r} = 
\overline{\la \hat\r_a \hat\r_b \ra_{\rm m}} = 
\overline{\la \hat\r_a\ra_{\rm m}\la \hat\r_b \ra_{\rm m}} =
\overline{\la \hat\r\ra_{\rm m}\la \hat\r \ra_{\rm m}}
\eeq
which is exactly the kind of average we want to compute.
Similarly, assuming that different letters denote different values of the indexes:
\beq\begin{split}
& \la \hat\r_a \hat\r_a\hat\r_b \ra_{\rm r} = 
\overline{\la \hat\r_a \hat\r_a\hat\r_b \ra_{\rm m}} = 
\overline{\la \hat\r_a\hat\r_a\ra_{\rm m}\la \hat\r_b \ra_{\rm m}} =
\overline{\la \hat\r\hat\r\ra_{\rm m}\la \hat\r \ra_{\rm m}} \ , \\
& \la \hat\r_a \hat\r_b\hat\r_c \ra_{\rm r} = 
\overline{\la \hat\r_a \hat\r_b\hat\r_c \ra_{\rm m}} = 
\overline{\la \hat\r_a \ra_{\rm m} \la\hat\r_b\ra_{\rm m}\la \hat\r_c \ra_{\rm m}} =
\overline{\la \hat\r \ra_{\rm m} \la\hat\r\ra_{\rm m}\la \hat\r \ra_{\rm m}}
\end{split}\eeq

Let us introduce a space-dependent order parameter
\beq 
\hat q_{ab}(r) =  \int \de x \, f(x) \, \hat\r^{(2)}_{ab}(r-x/2,r+x/2)  \ ,
\eeq
and the two-replica correlation function
\beq \label{def_order_parameter}
C_{ab}(r) = \la \hat q_{ab}(r) \ra_{\rm r} - \r^2 =  \int \de x ~ f(x) [ \langle \hat\r^{(2)}_{ab}(r-x/2,r+x/2) \rangle_{\rm r} - \langle \hat\r_a(r-x/2) \rangle_{\rm r} \langle \hat \r_b(r+x/2) \rangle_{\rm r} ]  \ ,
\eeq
where $f(x)$ is once again an arbitrary short ranged function.
We are interested in these functions for $a \neq b$. Using the prescriptions above we obtain
\beq
C_{ab}(r) = \int \de x \, f(x) \, [ \overline{\langle \hat\r_a(r-x/2) \rangle_{\rm m} \langle  \hat \r_b(r+x/2) \rangle_{\rm m}} ] - \r^2 \ .
\eeq
At this point the replica indexes can be dropped because the one-replica average in a metastable state is the same for all replicas, and we get
\beq
C_{ab}(r) = \int \de x \, f(x) \, [ \overline{\langle \hat\r(r-x/2) \rangle_{\rm m} \langle  \hat \r(r+x/2) \rangle_{\rm m}} - \r^2 = \mathbf E[\langle  \hat C(r,t \to\io) \rangle] - \r^2\ ,
\eeq
which provides the crucial identification between replicas and the long time limit of dynamics in a metastable state.

Similar mappings can be obtained for four-point correlations.
We define the correlation matrix of the order parameter as (for $a\neq b$ and $c\neq d$):
\beq\label{Gfdef}
G^{(f)}_{ab;cd}(r) = 
\la \hat q_{ab}(r) \hat q_{cd}(0) \ra_{\rm r} - \la \hat q_{ab}(r) \ra_{\rm r} \la \hat q_{cd}(0) \ra_{\rm r} \ ,
\eeq
where the superscript $f$ is useful to keep in mind that we performed a smoothing through the function $f(x)$.
Performing similar manipulations as for the two-point functions, we have
\beq
G^{(f)}_{ab;cd}(r) =  \int \de x \de y \, f(x) f(y) \, [ \overline{\langle \hat\r_a(r-x/2) \hat \r_b(r+x/2)
\hat\r_c(-y/2)  \hat \r_d(y/2) \rangle_{\rm m} } ] 
-\mathbf E[\langle  \hat C(r,t \to\io) \rangle] \mathbf E[\langle  \hat C(0,t \to\io) \rangle] 
\eeq
and in the first term the average $\la \bullet \ra_{\rm m}$ can be factorized over different indexes.
Comparing this with Eq.~(\ref{Gio}) we obtain:
\beq\label{GdGr}
\begin{split}
G_4(r,t\to\io) &= G^{(f)}_{ab;ab}(r) \ , \\
G_{th}(r,t\to\io)   &=G^{(f)}_{ab;ab}(r)-G^{(f)}_{ab;ac}(r) \ .
\end{split}\eeq

\subsection{The replicated free energy} 
\label{sec:defD}

We now discuss how replica correlation functions can be computed. We introduce some standard notations
of liquid theory~\cite{Hansen} and we adapt them to the replicated system. 

Let us start with the grand canonical partition function for a $D$-dimensional fluid with pairwise additive potential 
$v$, chemical potential $\m$, and
under an external field $\Psi$. The logarithm of the partition function reads:
\begin{equation}
W[\n,w] = \ln Z[\n,w] = \ln \sum_{N=0}^\infty
\frac{1}{N!}\int\left[\prod_{i=1}^N \de x_i\right] \exp \left( \frac 12 \int
  \de x \de y ~ \hat{\r}^{(2)}(x,y) w(x,y) + \int \de x ~ \n(x) \hat{\r}(x) \right)
\end{equation}
where we have used the following definitions for the fields
\beq
\begin{split}
& \hr(x) = \sum_{i=1}^N \d(x-x_i) \ , \\ 
& \hrd(x,y) = \sum_{i=1}^N \sum_{j \ne i} \d(x-x_i) \d(y-x_j) 
= \hr(x)\hr(y) - \hr(x) \delta(x-y)
\end{split}
\eeq
and the microscopic details of the system are encoded in
\beq
\begin{split}
& \n(x) = \b \m - \b \Psi(x)  \ , \\
& w(x,y) = - \b v(x,y) \ .
\end{split}
\eeq

To study the glassy phase we will follow the method introduced in \cite{Mo95,MP96}. We replicate the system introducing other $m-1$ copies of this original fluid, with interaction between copy $a$ and copy $b$ denoted by $w_{ab} = -\b v_{ab}$, so that the logarithm of the replicated partition function is given by
\begin{align}
& W[\{ \n_a \},\{ w_{ab}\}] = \ln Z[\{ \n_a \} ,\{ w_{ab} \} ] = \\
& =\ln \sum_{N=0}^\infty\frac{1}{(N!)^m}\int \left(\prod_{a=1}^m \prod_{i=1}^N \de x_i^{(a)}\right) \exp \left( \frac 12 \sum_{a,b}^{1,m} \int \de x \de y 
~ \hrd_{ab}(x,y) w_{ab}(x,y) + \sum_{a=1}^m \int \de x ~ \n_a(x) \hr_a(x) \right) , 
\end{align}
where the definition of the fields must be modified in order to take into account different replicas
\beq
\begin{split}
& \hr_a(x) = \sum_{i=1}^N \d(x - x_{i_a}) \ , \\
& \hrd_{ab}(x,y) = \hr_a(x) \hr_b(y) 
- \hr_a(x) \d_{ab} \d(x-y) \ .
\end{split}
\eeq
In the following, to lighten the notations, we will sometimes (when this leads to no ambiguity):
\begin{enumerate}
\item use shorthand
notations for the spatial positions, e.g. $f(1) \to f(x_1)$, $f(1,2) \to f(x_1,x_2)$;
\item
similarly, use $\int \de x_1 \to \int_1$;
\item
drop the replica and space indexes and simply denote 
$\hr \to \{\hr_a(1)\}$, $\hrd \to \{ \hrd_{ab}(1,2) \}$;
\end{enumerate}
and similarly for similar or more complex quantities.

To study the glassy phase we need to know the correlation functions of the two fields $\hr$ and $\hrd$. 
Let us underline that in this scheme, the details of the (attractive) interaction between different replicas is not important because in the end we will send it to zero. In fact if there is a glassy phase below at a certain dynamical temperature $T_d$, this infinitesimal attractive potential is enough to let all the $m$ replicas fall down in the same state. 
Hence, it is very convenient to perform a double Legendre transform and write the free energy as a function of the averages of $\hr$ and $\hrd$, which
we denote $\r$ and $\rd$~\cite{MH61,DM64,CJT74}.
In this way, we can take directly the limit where there is no interaction between different replicas and look for a solution where the replicas remain correlated in this limit.
We obtain~\cite{MP96}:
\begin{align}
& \G[\r, \rd] = \sum_a \int \r_a(1) \n_a^*(1) + \frac 12 \sum_{a,b} \int_{1,2}\r_{ab}^{(2)}(1,2) w_{ab}^*(1,2) 
- W[\n^*,w^* ] \ ,
\end{align}
where $\n^*$ and $w^*$ are the solution of the two equations
\begin{align}
& \left. \frac{\d W[\n,w]}{\d \n_a(1)} \right|_{\n^*,w^*} = \r_a(1) ~\text{ and }~ \left. \frac{\d W[\n,w]}{\d w_{ab}(1,2)} 
\right|_{\n^*,w^*} = \frac 12 \rd_{ab}(1,2) . \label{double_LT}
\end{align}
Morita and Hiroike~\cite{MH61} showed that this double Legendre transform can be written as 
\beq\label{GammaGen}
\G[\r,\rd] = \G_{\rm id}[\r,\rd] + \G_{\rm ring}[\r,\rd] + \G_{\rm 2PI}[\r,\rd] \ ,
\eeq
where
\beq\label{Gammaidring}
\begin{split}
\G_{\rm id}[\r,\rd] &=  \sum_a \int_1 \r_a(1) \left[ \ln \r_a(1) - 1 \right] + \frac 12 \sum_{a,b} \int_{1,2} \left[ \r^{(2)}_{a,b}(1,2) \ln \left( \frac{\r^{(2)}_{a,b}(1,2)}{\r_a(1) \r_b(2)}
 \right) -  \r^{(2)}_{a,b}(1,2) + \r_a(1) \r_b(2) \right]  \ , \\
\G_{\rm ring}[\r,\rd] &=  \frac 12 \sum_{n \ge 3} \frac{(-1)^n}{n} \sum_{a_1,\ldots,a_n} 
\int_{1,\ldots,n} \r_{a_1}(1) h_{a_1 a_2}(1,2) \cdots \r_{a_n}(n) h_{a_n a_1}(n,1) 
\end{split}
\eeq
and we have introduced
\beq
h_{ab}(1,2) = \frac{\r^{(2)}_{a,b}(1,2)}{\r_a(1) \r_b(2)} -1 \ .
\eeq 
The term $\G_{\rm 2PI}$ is the sum of all two-particle irreducible diagrams that are defined precisely in~\cite{MH61,DM64,CJT74}. We do not give more details
because this term will be mostly neglected when we will perform concrete numerical computations. However, the formalism we develop below holds in
full generality so one could include 2PI diagrams in future works. For instance, this can be done through a systematic expansion in powers of the
off-diagonal term of $\rd$. The price to pay is that the result depends on many-body correlations of the non-replicated liquid~\cite{JZ12}.

It will be useful in the following to
define the direct correlation function $c_{ab}(1,2)$ 
through a replicated version of the Ornstein-Zernike equation:
\begin{align}
& h_{ab}(1,2) = c_{ab}(1,2) + \sum_c \int_3 h_{ac}(1,3) \r_c(3) c_{cb}(3,2) , \label{OZ_rep}
\end{align}
whose solution can be written through a series expansion in the following way
\begin{align}
& c_{ab}(1,2) = \sum_{n \ne 1} (-1)^n \sum_{a_2,\ldots,a_{n-1}} \int_{3,\ldots,n-1} h_{a a_1}(1,3) 
\r_{a_1}(3) h_{a_1 a_2}(3,4) \cdots \r_{a_{n-1}}(n-1) h_{a_{n-1}b}(n-1,2) .
\end{align}
The free-energy is computed by evaluating the functional $\G$ at the physical correlator $\overline \r_{ab}(1,2)$ which solves the following equation \cite{MH61,DM64,CJT74}:
\begin{align}
\left. \frac{\d \G[\r,\r^{(2)}]}{\d \r^{(2)}_{ab}(1,2)} \right|_{\overline \r_{ab}} = \frac 12 w_{ab}(1,2) 
\label{extremum_g} .
\end{align}
The density field can be determined by a similar equations as a function of the chemical potential, however here we are interested in
a solution with $\r_a(1)= \r$, hence we can directly fix the density in this way.
Moreover, we are interested in a solution for
the two point function which has eventually (below the dynamical temperature $T_d$) a 1RSB structure
\beq\label{SP}
\overline \rd_{ab}(1,2) = \d_{ab} \r^2 g(1,2) + (1-\d_{ab}) \r^2 \wt g(1,2)
\eeq
In particular in the high temperature phase we expect that the off-diagonal part of this solution, namely $\tilde g$, is trivial (it corresponds to uncorrelated replicas, hence
it is identically equal to 1) while below the dynamical temperature $T_d$ we have a non trivial solution. 
Note that the glass transition can be crossed either by lowering temperature or by increasing density, the second strategy being the only possible one for hard spheres
like systems. In the following general discussion we will typically refer to lowering temperature: but all of our results apply to any other path in the phase diagram that crosses the glass
transition line. We will indeed present concrete numerical calculations both in temperature and density.

\section{Landau expansion of the free energy around the glassy solution}

In this section, we will show how one can construct a Landau expansion of the free energy.
There are two main differences with respect to the simple ferromagnetic example of Sec.~\ref{sec:ill}.
First of all, even for a homogeneous system, the order parameter is $\wt g(x_1-x_2)$ and it keeps a non trivial dependence on space.
Second, at the mean field level the glass transition is a {\it random first order transition}: even if it is a second order transition from a thermodynamic
point of view, the order parameter has a finite jump at the critical (dynamical) temperature $T_d$. This implies that we cannot approach smoothly
the glass phase from the liquid one. In the following, we will assume that we are in the glass phase at $T<T_d$, where $\wt g$ is non-trivial, and study
how the limit of $\ee = T_d - T \to 0$ is approached from positive $\ee$.

In Sec.~\ref{sec:LandauA} we introduce an appropriate scalar order parameter and perform a Landau expansion of the free energy for small deviations
of the order parameter around the critical point. In Sec.~\ref{sec:LandauB} we show how this expansion can be used to compute the MCT exponents,
in particular the exponent $\l$, following~\cite{CFLPRR12}.
In Sec.~\ref{sec:LandauC} we perform a more detailed study of the mass matrix, i.e. the quadratic term of the expansion.
In Sec.~\ref{sec:LandauD} we use this to show that the value of the MCT critical exponents do not depend on the details of the definition of the scalar
order parameter; as a side product we obtain a much simpler expression for these exponents.

\subsection{Free energy for a uniform field}
\label{sec:LandauA}

%
We want to define the free energy as a function of the order parameter $C_{ab}(r)$, defined in Eq.~\eqref{def_order_parameter}, 
in the case in which it is uniform, i.e. independent of $r$. The way we can produce this quantity is just by 
maximizing the free energy with respect to $\rd$ under the constraint that $C_{ab}$ is given by its definition, which is enforced through a Lagrange
multiplier $\e_{ab}$:
\beq
\G[C_{ab}] = \max_{\rd, \, \e_{ab}} \left[ \G[\r,\rd] - \sum_{a\neq b} \e_{ab} \left( V C_{ab} -  \int_{1,2} f(1,2) [ \rd_{ab}(1,2) - \r_a(1) \r_b(2) ] \right) \right] \ .
\eeq
We now that in absence of the constraint, $\e_{ab}=0$, the free energy $\G[\r,\rd]$ is maximum for $\rd = \overline{\rd}$, Eq.~\eqref{SP}, which corresponds to some value $\overline{C}_{ab}$
of the order parameter. 
We want to expand around this reference solution.
We call $\D \rd = \rd - \overline{\rd}$ and $\D C_{ab} = C_{ab} - \overline{C}_{ab}$ the deviations from this reference solution. 

We use from now on a lighter notation in which a global index is used instead of replica and space indices, $A = \{a,b,1,2\}$, $B = \{c,d,3,4\}$; moreover we use the Einstein's convention where repeated indices are implicitly summed. 
Expanding $\G[\r,\rd]$ around the reference solution we obtain (recall that $\r = \r_a(1) = \r_b(2)$ is a fixed constant)
\beq\begin{split}
\D\G[\D C_{ab}] = \max_{\D\r^{(2)} \, , \ \e_{ab}} & \left\{
\frac12 \frac{\d^2\G[\r,\overline{\rd}]}{ \d \r^{(2)}_A \d\r^{(2)}_B } \D\rd_A \D\rd_B 
+ \frac16 \frac{\d^3\G[\r,\overline{\rd}]}{ \d \r^{(2)}_A \d\r^{(2)}_B \d\r^{(2)}_C } \D\rd_A \D\rd_B \D\rd_C \right. \\
& \left. - \sum_{a \neq b} \e_{ab} \left( V \D C_{ab} -  \int_{1,2} f(1,2) \D \rd_{ab}(1,2)   \right) \right\} \ ,
\end{split}\eeq
where all the derivatives must be computed in the reference solution. Defining 
\beq\begin{split}\label{mass_and_cubic_interactions}
& M_{AB} =  \frac{\d^2\G[\r,\overline{\rd}]}{ \d \r^{(2)}_A \d\r^{(2)}_B } \ , \\
& L_{ABC} = \frac{\d^3\G[\r,\overline{\rd}]}{ \d \r^{(2)}_A \d\r^{(2)}_B \d\r^{(2)}_C } \ , \\
\end{split}\eeq
and $\e_A = \e_{ab} f(1,2)$ (where we define $\e_{aa}=0$),
the derivative with respect to $ \D\r^{(2)}$ leads to the following equation:
\beq
0 = M_{AB} \D\rd_B + \frac12 L_{ABC} \D\rd_B \D\rd_C + \e_A \ ,
\eeq
which can be inverted perturbatively and gives
\beq
\D \rd_A = - M^{-1}_{AB} \e_B - \frac12 M^{-1}_{AB} L_{BCD} M^{-1}_{CC'} \e_{C'} M^{-1}_{DD'} \e_{D'} + O(\e^3)  \ .
\eeq
Plugging this in the free energy we obtain:
\beq\label{appDG}
\begin{split}
\D\G[\D C_{ab}] = \max_{ \e_{ab}} & \left[  
-\frac12 \e_A M^{-1}_{AB} \e_B - \frac16   L_{ABC} M^{-1}_{AA'} M^{-1}_{BB'} M^{-1}_{CC'}  \e_{A'} \e_{B'} \e_{C'} 
 - V\sum_{a \neq b} \e_{ab}  \D C_{ab} \right] \ .
\end{split}\eeq
At this point it is convenient to recall that $\e_A = \e_{ab} f(1,2)$. 
Using the shorthand notation $(M^{-1} M^{-1} M^{-1} L)_{ABC} = M^{-1}_{AA'} M^{-1}_{BB'} M^{-1}_{CC'} L_{A'B'C'}$
and introducing
\beq\label{MMLLdef}
\begin{split}
&V \MM^{-1}_{ab,cd} = \int_{1,2,3,4} f(1,2) M^{-1}_{ab,cd}(1,2,3,4) f(3,4) \ , \\
&V \LL_{ab,cd,ef} = \int_{1,\cdots,6} f(1,2) f(3,4) f(5,6) (M^{-1} M^{-1} M^{-1} L)_{ab,cd,ef}(1,2;3,4;5,6) 
\end{split}\eeq
we can rewrite Eq.~(\ref{appDG}) as
\beq\label{app2DG}
\begin{split}
\D\G[\D C_{ab}] = V \max_{ \e_{ab}} & \left[  
-\frac12 \e_{ab} \MM^{-1}_{ab,cd} \e_{cd} - \frac16   \LL_{ab,cd,ef}  \e_{ab} \e_{cd} \e_{ef} 
 - \sum_{a \neq b} \e_{ab}  \D C_{ab} \right] \ .
\end{split}\eeq
We now take the derivative with respect to $\e_{ab}$ and we obtain
\beq
\D C_{ab} = - \MM^{-1}_{ab,cd} \e_{cd} - \frac12 \LL_{ab,cd,ef} \e_{cd} \e_{ef} \ ,
\eeq
which is inverted as
\beq
\e_{ab} = -\MM_{ab,cd} \D C_{cd} -\frac12 \MM_{ab,cd} \LL_{cd,ef,gh}  \MM_{ef,e'f'} \D C_{e'f'}  \MM_{gh,g'h'} \D C_{g'h'}  \ ,
\eeq
and plugging this in Eq.~\eqref{app2DG} 
we finally obtain the desired third order Landau expansion of the free energy, which is the analog of Eq.~\eqref{Landau_ferro} for the glass
transition:
\beq\label{eq:gamma_delta_q}
\begin{split}
\D\G[\D C_{ab}] &= V \left\{
\frac12 \D C_{ab} \MM_{ab,cd} \D C_{cd} + \frac16 \WW_{ab,cd,ef} \D C_{ab} \D C_{cd} \D C_{ef} + \cdots \right\} \ , \\
\WW_{ab,cd,ef} &=   \MM_{ab,a'b'} \MM_{cd,c'd'} \MM_{ef,e'f'} \LL_{a'b',c'd',e'f'} \ .
\end{split}\eeq

\subsection{Computation of $\l$}
\label{sec:LandauB}

At this point we are equipped with all the ingredients
to give an explicit expression for the parameter exponent $\l$ of MCT. 
It has been shown in \cite{CFLPRR12} that in mean field disordered systems this parameter can be computed in a purely static framework. 
The argument is the following: the dynamics of such systems can be studied in great detail using the Martin-Siggia-Rose \cite{MSR73} formalism; 
in particular, by going to the supersymmetric representation for the action of the Langevin process describing such dynamics \cite{Ku92}, 
one can obtain a dynamical Gibbs free energy which has exactly the same form (apart from a kinetic term containing the derivatives with respect 
to time which can be neglected if we want to study the long-time behavior) of the one in (\ref{eq:gamma_delta_q}) where replica indices are replaced by supertimes. 
From the dynamical action it is straightforward to see that the exponent parameter $\l$ is given by the ratio between two of the cubic terms of the 
dynamical Gibbs free energy. But now the key point is that this two coefficients are related to six-points dynamical correlation functions whose 
long time behavior can be studied using replicas following the same analysis of Sec.~\ref{sec:defC}. 
In particular it can be shown that for times such that the two point correlation function is very close 
to its plateau value, the values of these dynamical six-points correlation functions can be computed in a static framework just by using the replicated 
Gibbs free energy of the form (\ref{eq:gamma_delta_q}). This implies directly that the exponent parameter can be computed from the statics. 
This argument has been discussed in full detail in~\cite{PR12}.

The cubic coefficients of the replicated Gibbs free energy which are relevant for the computation of the $\l$ are the following ones
\beq
-\frac{w_1}{6}\Tr (\Delta C)^3-\frac{w_2}{6}\sum_{a\neq b}\Delta C_{ab}^3\:.
\eeq
and the exponent parameter is given by
\beq
\l=\frac{w_2}{w_1}\:.
\eeq
In computing the derivatives in (\ref{mass_and_cubic_interactions}) one has to use the explicit form for the order parameter $\overline{\rd}$ 
that maximizes the free energy, given in Eq.~\eqref{SP}, which is replica symmetric (corresponding in our formalism to a 1RSB solution).
Exploiting this symmetry, in \cite{TDP02} it has been shown that for a replica symmetric saddle point the two coefficients $w_1$ and $w_2$ can be written in the following form
\beq\label{DeDominicis}
\begin{split}
w_1&=\WW_1-3\WW_5+3\WW_7-\WW_8 \ , \\
w_2&=\frac{1}{2}\WW_2-3\WW_3+\frac{3}{2}\WW_4+3\WW_5+2\WW_6-6\WW_7+2\WW_8 \ ,
\end{split}
\eeq
where
\beq\label{DeDominicis2}
\begin{split}
&\WW_1=\WW_{ab,bc,ca} \ , \\
&\WW_2=\WW_{ab,ab,ab}\ ,\\
&\WW_3=\WW_{ab,ab,ac}\ ,\\
&\WW_4=\WW_{ab,ab,cd}\ ,\\
&\WW_5=\WW_{ab,ac,bd}\ ,\\
&\WW_6=\WW_{ab,ac,ad}\ ,\\
&\WW_7=\WW_{ac,bc,de}\ ,\\
&\WW_8=\WW_{ab,cd,ef} \ .
\end{split}
\eeq
Therefore the final expression for $\l$ is
\beq
\l = \frac{w_2}{w_1} = \frac{\frac{1}{2}\WW_2-3\WW_3+\frac{3}{2}\WW_4+3\WW_5+2\WW_6-6\WW_7+2\WW_8}{\WW_1-3\WW_5+3\WW_7-\WW_8}\:.	
\eeq
Note that this expression contains implicitly a dependence on the function $f(x)$ that we have chosen to define the order parameter in Eq.~\eqref{def_order_parameter}. 
Moreover, note that the usefulness of this relation is very hard to prove at this stage because to compute the cubic coefficients $\WW$ 
we have to perform a complex operator inversion
to obtain the $\MM$, see Eq.~\eqref{MMLLdef}, followed by a convolution with the $\LL$, see Eq.~\eqref{eq:gamma_delta_q}. This would be an extremely hard numerical
calculation.
In fact, all these problems can be solved noting that the mass matrix of the free energy develops a zero mode at the dynamical transition, and the presence of 
this zero mode simplifies a lot the computation of $\l$, as we show next.

\subsection{The mass matrix and the zero mode}
\label{sec:LandauC}

Recall that here we are looking to a homogeneous system. Hence, in Eq.~(\ref{SP}) we
have $\tilde g(x_1,x_2) = \tilde g(x_1-x_2)$. Keeping this in mind, in the following
we will use equivalently the notations $\tilde g(x_1,x_2)$, $\tilde g(1,2)$, and $\tilde g(x)$.
As we already explained, at the mean field level
the solution of the saddle point Eq.~(\ref{SP}) is $\tilde g(x)=1$ in the liquid phase while
$\tilde g(x)$ is a non-trivial function in the glass phase.
The appearance of an off-diagonal solution below the dynamical transition point is 
discontinuous and can be regarded as a bifurcation-like phenomenon at some critical 
value of the control parameters (temperature or density). 
Because the transition is discontinuous, the bifurcation does not happen around the liquid
solution $\tilde g(x)=1$, but around a non trivial value of $\tilde g(x)$ corresponding
to the solution at the critical point. Therefore, if we approach the transition point
from the liquid phase and look at the behavior of the free energy around the liquid solution,
nothing special will happen.
For this reason we are forced to take the control parameter such that
we are in the glass phase (for example we put $T<T_d$) and approach the transition from this phase.
Let us call $\ee=T_d-T$ the distance from the critical point. Then for $\ee\to 0^+$
\beq\label{sqrtSP}
\wt g(1,2;\ee) = \wt g(1,2; 0) + 2 \, \sqrt{\ee} \, \k \, k_0(1,2) + O(\ee) \ .
\eeq
Here, the function $k_0$ is normalized by $V^{-1}\int_{1,2}k_0(1,2)^2 = \int \de x k_0(x)^2=1$,
which defines implicitly the constant $\k$.
Therefore
\beq\label{zm2}
\frac{d \overline \r_{ab}(1,2)}{d\ee} = \frac{\r^2}{\sqrt{\ee}} (1-\d_{ab}) \, \k \, k_0(1,2) + O(1) \ .
\eeq
We consider the saddle point condition (\ref{extremum_g}) with $a \neq b$ and $w_{a\neq b}=0$. Then
\beq\label{app2der}
\begin{split}
0 &= \frac{d}{d\ee} \left. \frac{\d \G[\r,\r^{(2)}]}{\d \r^{(2)}_{ab}(1,2)} \right|_{\wt\r_{ab}}
= \sum_{c \neq d} \int_{3,4} \left. \frac{\d^2 \G[\r,\r^{(2)}]}{\d \r^{(2)}_{ab}(1,2)\d \r^{(2)}_{cd}(3,4)} \right|_{\overline\r_{ab}} \frac{d \overline\r_{cd}(3,4)}{d\ee} \\
&+ \sum_c  \int_{3,4} \left. \frac{\d^2 \G[\r,\r^{(2)}]}{\d \r^{(2)}_{ab}(1,2)\d \r^{(2)}_{cc}(3,4)} \right|_{\overline\r_{ab}} \frac{d \overline\r_{cc}(3,4)}{d\ee} + \left. \frac{\d^2 \G[\r,\r^{(2)}]}{\d \r^{(2)}_{ab}(1,2)\d \ee} \right|_{\overline\r_{ab}}
\end{split}\eeq
Recall the definition of the ``mass operator'' in Eq.~\eqref{mass_and_cubic_interactions} 
\beq\label{mass_def2}
M_{ab;cd}(1,2;3,4) =  \left. \frac{\d^2 \G[\r,\r^{(2)}]}{\d \r^{(2)}_{ab}(1,2)\d \r^{(2)}_{cd}(3,4)} \right|_{\overline\r_{ab}}
\eeq 
for $a\neq b$ and $c\neq d$. Then
we can write Eq.~\eqref{app2der} as
\beq\label{zm1}
\begin{split}
0 = \sum_{c\neq d} \int_{3,4} M_{ab;cd}(1,2;3,4) \frac{d \overline\r_{cd}(3,4)}{d\ee}  + \KK(1,2)
\end{split}\eeq
where $\KK(1,2)$ is finite at the critical point and does not depend on $a \neq b$ because of the symmetry of the saddle point.
Recalling from Eq.~\eqref{zm2} that the derivative of the saddle point solution is divergent 
when we approach the dynamical point from below, we conclude that the mass operator 
should develop a zero mode at the transiton, in such a way that the divergent part of Eq.~\eqref{zm1}
is cancelled. 
In other words, we should have that
\beq\label{zeroabcd}
\sum_{c\neq d}\int _{3,4}M_{ab;cd}(1,2;3,4)k_0(3,4)=\mu\sqrt\epsilon k_0(1,2) \ .
\eeq
Because of the replica symmetry of the saddle point solution, 
the most general form for this mass matrix is~\cite{TDP02}
\beq\label{mass}
M_{ab;cd}(1,2;3,4)=M_1(1,2;3,4)\left(\frac{\delta_{ac}\delta_{bd}+\delta_{ad}\delta_{bc}}{2}\right)+M_2(1,2;3,4)\left(\frac{\delta_{ac}+\delta_{ad}+\delta_{bc}+\delta_{bd}}{4}\right)+M_3(1,2;3,4)
\eeq
Because such matrices make a closed algebra,
the inverse of the mass operator must have the same replica structure:
\beq\label{inv_mass}
G_{ab;cd}(1,2;3,4)=G_1(1,2;3,4)\left(\frac{\delta_{ac}\delta_{bd}+\delta_{ad}\delta_{bc}}{2}\right)+ G_2(1,2;3,4)\left(\frac{\delta_{ac}+\delta_{ad}+\delta_{bc}+\delta_{bd}}{4}\right)+G_3(1,2;3,4) \ .
\eeq
The equation for $G$ is
\beq\label{invMG}
\sum_{c\neq d}\int_{3,4}M_{ab;cd}(1,2;3,4)G_{cd;ef}(3,4;5,6)=\frac{\delta_{ae}\delta_{bf}+\delta_{af}\delta_{be}}{2}\frac{\delta(1,5)\delta(2,6)+\delta(1,6)\delta(2,5)}{2}
\eeq
and, in the replica limit $m\to 1$, 
the solution of this equation is given 
by (we denote with $\otimes$ the convolution in the space variables $a\otimes b=\int_{3,4}a(1,2,;3,4)b(3,4;5,6)$):
\beq\label{Gelements}
\begin{split}
& G_1=M^{-1}_1\\
& G_2=-2[2M_1-M_2]^{-1}\otimes M_2\otimes M_1^{-1}\\
& G_3=M_1^{-1}\otimes\left\{ M_2\otimes[2M_1-M_2]^{-1}\otimes M_2-M_3\right\}\otimes M_1^{-1}\:.
\end{split}\eeq

We know from Eq.~\eqref{zm2} that the zero mode is independent of the off-diagonal replica indexes.
This implies for a matrix of the type (\ref{mass}) that in the replica limit $m\to 1$
\beq\begin{split}
& \lim_{m\to 1}\sum_{c\neq d}\int_{3,4}M_{ab;cd}(1,2;3,4)k_0(3,4)=\\
&=\lim_{m\to 1}\int_{3,4}\left[m(m-1)M_3(1,2;3,4)+M_1(1,2;3,4)+(m-1)M_2(1,2;3,4)\right]k_0(3,4) \\
&= \int_{3,4}M_1(1,2;3,4)k_0(3,4)
\end{split}\eeq
This implies that for $m=1$ 
among all the components of the mass operator, $k_0$ is a zero mode (with eigenvalue proportional to $\sqrt\epsilon$)
only for the operator defined by the kernel $M_1(1,2;3,4)$, 
while $M_2$ and $M_3$ do not need to have any zero mode at the transition. 
In formulae, from Eq.~\eqref{zeroabcd},
\beq\label{M1k0eq}
 \int_{3,4}M_1(1,2;3,4)k_0(3,4) = \mu\sqrt\epsilon k_0(1,2) \ ,
\eeq
or equivalently (recall that $\int_{1,2}k_0(1,2)^2 = V$):
\beq
M_1^{-1}(1,2;3,4) = \frac{1}{V \, \mu\sqrt\epsilon}  k_0(1,2)  k_0(3,4) + O(1) \ .
\eeq
This observation is very important because it shows that the operators defined by the kernels $G_1$ and $G_2$ 
have a single pole (a divergent eigenvalue) while the operator $G_3$ has a double pole. 
This is a very straightforward generalization of results obtained in~\cite{FPRR11} to the case in which the 
system has a non trivial spatial structure. Let us now rewrite Eq.~\eqref{Gelements} as
\begin{equation}
\begin{array}{ll}
 G_2=O_2\otimes M_1^{-1} & \text{ with } \ \ \ O_2 = -2[2M_1-M_2]^{-1}\otimes M_2 \\
G_3=M_1^{-1}\otimes O_3\otimes M_1^{-1} &  \text{ with } \ \ \  O_3 = \left[M_2\otimes[2M_1-M_2]^{-1}\otimes M_2-M_3\right] \\
\end{array}
\end{equation}
and use this to conclude that the most divergent contributions to these kernel operators are given by
\beq\label{Gleading}
\begin{array}{ll}
G_1(1,2;3,4)\simeq \frac{1}{V \, \mu\sqrt\epsilon}k_0(1,2)k_0(3,4)+O(1) &  \\ 
G_2(1,2;3,4)\simeq \frac{1}{V \, \mu\sqrt\epsilon}k_2(1,2)k_0(3,4)+O(1) & \text{ with } \ \ k_2(1,2)=\int_{3,4}O_2(1,2;3,4)k_0(3,4) \\ 
G_3(1,2;3,4)\simeq \frac{\k_3}{V \, \mu^2\epsilon}k_0(1,2)k_0(3,4)+O\hskip-3pt\left(\frac{1}{\sqrt\epsilon}\right) & \text{ with } \ \
  \k_3=\frac1V \int_{1,2,3,4}k_0(1,2)O_3(1,2;3,4)k_0(3,4) \\
\end{array}
\eeq
Note that in $G_3$ a term proportional to $1/\sqrt\epsilon$ appears and it depends on the exicited states of the kernel operator $M_1$. However we will see that the contribution of the zero mode is enough for the computation of $\lambda$.

\subsection{Analysis of the cubic terms and computation of $\l$}
\label{sec:LandauD}

At this point we are equipped to extract the divergent part of the cumulants $w_1$ and $w_2$ and obtain a simple and universal expression of $\l$.
Let us start with the generic expression (\ref{eq:gamma_delta_q}) for the cubic coefficients of the free energy as a 
function of the a uniform order parameter. Using Eq.~\eqref{MMLLdef} we have
\beq\label{Wapp}
\begin{split}
\WW_{ab,cd,ef} &= \MM_{ab;a'b'}\MM_{cd;c'd'}\MM_{ef;e'f'} \frac1V\int_{1,\ldots,6;1',\ldots,6'}f(1,2)M^{-1}_{a'b';a''b''}(1,2;1',2')\times \\
&\times f(3,4)M^{-1}_{c'd';c''d''}(3,4;3',4')f(5,6)M^{-1}_{e'f';e''f''}(5,6;5',6')L_{a''b'';c''d'';e''f''}(1',2';3',4';5',6') \\
&= \frac1V\int_{1',\ldots,6'} \D_{ab;a''b''}(1',2') \D_{cd;c''d''}(3',4')
\D_{ef;e'',f''}(5',6')L_{a''b'';c''d'';e''f''}(1',2';3',4';5',6')
\ ,
\end{split}\eeq
where we introduced the quantity
\beq\begin{split}
\D_{ab;a''b''}(1',2') &= 
 \MM_{ab;a'b'}\int_{1,2} f(1,2)M^{-1}_{a'b';a''b''}(1,2;1',2') \\
&=
\Delta_1(1',2')\frac{\delta_{a,a''}\delta_{b,b''}+\delta_{a,b''}\delta_{b,a''}}2+
\Delta_2(1',2') \frac{\delta_{a,a''}+\delta_{b,a''}+\delta_{a,b''}+\delta_{b,b''}}4+
 \Delta_3(1',2') \ .
\end{split}\eeq
Clearly the replica structure of the matrix $\MM_{ab;cd}$ is inherited from the structure of the mass matrix $M_{ab;cd}(1,2;3,4)$ 
and its inverse. From Eq.~\eqref{MMLLdef} we have
\beq\begin{split}
\MM^{-1}_{ab,cd} &=\GG_1\left(\frac{\delta_{ac}\delta_{bd}+\delta_{ad}\delta_{bc}}{2}\right)+\GG_2\left(\frac{\delta_{ac}+\delta_{ad}+\delta_{bc}+\delta_{bd}}{4}\right)+ \GG_3 \ , \\
\MM_{ab,cd} &=\MM_1\left(\frac{\delta_{ac}\delta_{bd}+\delta_{ad}\delta_{bc}}{2}\right)+\MM_2\left(\frac{\delta_{ac}+\delta_{ad}+\delta_{bc}+\delta_{bd}}{4}\right)+ \MM_3 \ ,
\end{split}\eeq
 where we have defined the following $f$-dependent quantities
\beq
\GG_i=\frac1V \int_{1,2,3,4}f(1,2)G_i(1,2;3,4)f(3,4) \ ,
\eeq
and where using the same algebraic structure of the replica sector
\beq\begin{split}
&\MM_1=\frac{1}{\GG_1} \ ,\\
&\MM_2=-\frac{2\GG_2}{\GG_1(2\GG_1-\GG_2)} \ ,\\
&\MM_3=\frac{\GG_2^2}{\GG_1^2(2\GG_1-\GG_2)}-\frac{\GG_3}{\GG_1^2} \ .
\end{split}\eeq
Next we have to analyze the matrix $\D$ that appears in Eq.~\eqref{Wapp};
performing the matrix multiplications we obtain
\beq\begin{split}
&\Delta_1(1',2')=f \star \frac{1}{\GG_1}G_1 \ , \\
&\Delta_2(1',2')=f \star \frac{2(\GG_1G_2-G_1\GG_2)}{\GG_1(2\GG_1-\GG_2)} \ , \\
&\Delta_3(1',2')=f \star \frac{\GG_2^2G_1+2\GG_1(\GG_1G_3-\GG_3G_1)+\GG_2(\GG_3G_1-\GG_1(G_2+G_3))}{\GG_1^2(2\GG_1-\GG_2)} \ ,
\end{split}\eeq
where we used the shorthand notation $(f \star G_i)(1',2') = \int_{1,2} f(1,2) G_i(1,2;1',2')$.
Now we can use the critical behaviour of $G_i$ in Eq.~\eqref{Gleading}. We see that divergent terms of the same order will appear
at the numerator and denominator of the above expressions, so we can extract the finite part of $\D_i$ in the limit $\ee\to 0^+$.
Thanks to some non-trivial cancellations 
the result is, defining $f\star k_0=V^{-1} \int_{1,2} f(1,2)k_0(1,2) = \int \de x f(x) k_0(x)$:
\beq\begin{split}
&\Delta_1(1,2)= \frac{k_0(1,2)}{f\star k_0} \ , \\
&\Delta_2(1,2)=0 \ ,\\
&\Delta_3(1,2)=0 \ ,
\end{split}\eeq
and plugging this in Eq.~\eqref{Wapp} the final expression for the cubic coeffcients is given by
\beq
\WW_{ab;cd;ef}=\frac{1}{(f\star k_0)^3} \frac1V \int_{1,2;3,4;5,6}k_0(1,2)k_0(3,4)k_0(5,6)L_{ab;cd;ef}(1,2;3,4;5,6) \ .
\eeq
Using Eq.~\eqref{DeDominicis} we obtain the two relevant coefficients for the computation of $\l$ as:
\beq \label{w1_w2}
\begin{split}
w_2 &=\frac{1}{(f\star k_0)^3}\frac1V\int_{1,2;3,4;5,6}k_0(1,2)k_0(3,4)k_0(5,6)\Big(\frac{1}{2}L_{ab,ab,ab}-3L_{ab,ab,ac}+\frac{3}{2}L_{ab,ab,cd}+3L_{ab,ac,bd}+2L_{ab,ac,ad}\\
&-6L_{ac,bc,de}+2L_{ab,cd,ef}\Big)(1,2;3,4;5,6)\\
w_1&=\frac{1}{(f\star k_0)^3}\frac1V\int_{1,2;3,4;5,6}k_0(1,2)k_0(3,4)k_0(5,6)(L_{ab,bc,ca}-3L_{ab,ac,bd}+3L_{ac,bc,de}-L_{ab,cd,ef})(1,2;3,4;5,6)\:.
\end{split}\eeq
Note that because $\l$ is the ratio between $w_2$ and $w_1$,
the dependence on $f$ disappears and we get a consistent $f$-independent expression for the exponent parameter.
Moreover, the expression above are much simpler to compute because they simply require a convolution of the functions $L$ with the
zero mode $k_0$, without any need for operator inversion.

\section{Gradient expansion}

We want now to study the critical behavior of $G_{th}(r)$ and $G_4(r)$. 
These quantities are related by Eq.~\eqref{GdGr} to the inverse of the mass operator $G_{ab;cd}(1,2;3,4)$,
Eqs.~\eqref{mass} and \eqref{inv_mass}. 
In principle we should invert the mass operator $M_{ab;cd}(1,2;3,4)$ and perform a convolution with the smoothing
function $f(x)$ to obtain the replica correlation $G^{(f)}_{ab;cd}(r)$ that enters in Eq.~\eqref{GdGr}.
Although certainly well defined, this procedure would be numerically quite heavy even in the simplest approximation
for the mass matrix (e.g. the HNC approximation we discuss below).

In this section we want to show that, if we are only interested in the long distance behaviour of $G^{(f)}_{ab;cd}(r)$,
we can take advantage from the existence of the zero mode discussed in Sec.~\ref{sec:LandauC} to obtain a universal critical
form of $G^{(f)}_{ab;cd}(r)$. In particular we will show that the function $f(x)$ only enters in the prefactor, while the correlation
length is independent of $f(x)$.
For this we need to set up a gradient expansion where we consider a field $\D\rd_{ab}(1,2)$ which is almost uniform.
Recall that in the uniform case $\D\rd_{ab}(x_1,x_2) = \D\rd_{ab}(x_1-x_2)$. Here we consider a non-uniform field, that depends also
on the variable $(x_1+x_2)/2$, but we consider that the dependence on this variable is weak.

\subsection{Fourier transform}
\label{sec:gradA}

It is convenient to separate the dependence on the translationally invariant variable $x_1-x_2$ from the slow dependence
on space $(x_1+x_2)/2$ by introducing a Fourier transform
\beq\begin{split}
&\Delta \rd_{ab}(p,q) =\int \de x_1 \de x_2 \,
e^{i p \left(\frac{x_1+x_2}{2}\right) + iq \left(x_1 - x_2\right)}\Delta \rd_{ab}(x_1,x_2) \ , \\
&\Delta \rd_{ab}(x_1,x_2) =\int \frac{\de p \, \de q}{(2\pi)^{2D}} \,
e^{-i p \left(\frac{x_1+x_2}{2}\right) - iq \left(x_1 - x_2\right)}\Delta \rd_{ab}(p,q) \ .
\end{split}\eeq
Here $p$ is the momentum coupled to the slow spatial variation, while $q$ is the momentum coupled to local displacement.
Hence we will be interested in the limit of small $p$.
Plugging this in the quadratic part of the expansion of the free energy around the saddle point solution, we obtain
\beq\label{D2Ggrad}
\begin{split}
\Delta_2\Gamma &= \frac{1}{2}\sum_{a\neq b\ c\neq d}\int_{1,2,3,4}\Delta\rd_{ab}(1,2)M_{ab;cd}(1,2,3,4)\Delta\rd_{cd}(3,4) \\
&=\frac{1}{2}\sum_{a\neq b\ c\neq d}\int \frac{\de p \, \de \hat p \, \de q \, \de k}{(2\pi)^{4D}}
\Delta\rd_{ab}(p,q)M_{ab;cd}(-p,-\hat p;-q,-k)\Delta\rd_{cd}(\hat p,k) \ .
\end{split}
\eeq
where
\beq
M_{ab;cd}(p,\hat p;q,k) = \int_{1,2,3,4}
e^{ i p \left(\frac{x_1+x_2}{2}\right) +iq \left(x_1 - x_2\right)
+i \hat p \left(\frac{x_3+x_4}{2}\right) + ik \left(x_3 - x_4\right) }
M_{ab;cd}(1,2;3,4)
\eeq
Because of the translational invariance of the saddle point solution, the mass matrix is also translational invariant, so that
we can make a change of variable to $X = (x_1+x_2+x_3+x_4)/4$ and $u_i = x_i -X$, with $\sum_{i=1}^4 u_i =0$, and then
$M(1,2;3,4)$ does not depend on $X$. 
Calling $\DD u =\de u_1 \de u_2 \de u_3 \de u_4 \d\left(\frac14\sum_{i=1}^4 u_i\right)$ we get
\beq
\begin{split}
M_{ab;cd}(p,\hat p;q,k) & =  \int \de X \DD u \,
e^{i(p+\hat p) X + i p \left(\frac{u_1+u_2}{2}\right) + iq \left(u_1 - u_2\right)
+i \hat p \left(\frac{u_3+u_4}{2}\right) + ik \left(u_3 - u_4\right) }
M_{ab;cd}(u_1,u_2;u_3,u_4) \\
& = (2\pi)^D\delta(p+\hat p)M_{ab;cd}^{(p)}(q,k) \ , \\
M_{ab;cd}^{(p)}(q,k) &= \int \DD u \,
e^{ i p \left(\frac{u_1+u_2}{2}-\frac{u_3+u_4}{2}\right) +iq \left(u_1 - u_2\right) + ik \left(u_3 - u_4\right) }
M_{ab;cd}(u_1,u_2;u_3,u_4) \ .
\end{split}
\eeq

We want to study the correlation function	
\beq\begin{split}
G^{(p)}_{ab;cd}(q,k) 
&=\int \DD u \, e^{ip\left(\frac{u_1+u_2}{2}-\frac{u_3+u_4}{2}\right)+iq(u_1-u_2)+ik(u_3-u_4)}
\langle\Delta\hrd_{ab}(u_1,u_2)\Delta\hrd_{cd}(u_3,u_4)\rangle_{\rm r}  \\
&=\int \DD u \, e^{ip\left(\frac{u_1+u_2}{2}-\frac{u_3+u_4}{2}\right)+iq(u_1-u_2)+ik(u_3-u_4)}
G_{ab;cd}(u_1,u_2;u_3,u_4)
\ ,
\end{split}\eeq
when the two points $u_1,\ u_2$ are far away from $u_3,\ u_4$.
This implies that as already announced we need to study the limit $p\to 0$ and we have to develop the mass matrix around this limit. 
The correlation function above, at the Gaussian level, is given by the inverse of the mass matrix,
i.e. by transforming Eq.~\eqref{invMG} to Fourier space we get
\beq
\sum_{c\neq d}\int \frac{\de k}{(2\pi)^D} M_{ab;cd}^{(p)}(q,k) G_{cd;ef}^{(p)}(-k,q')
=\frac{\delta_{ae}\delta_{bf}+\delta_{af}\delta_{be}}{2} \, (2\pi)^D\frac{\delta(q-q')+\d(q+q')}2 \ .
\eeq
Therefore it has the form, akin to Eq.~\eqref{Gelements},
\beq\label{Gpqk}
\begin{split}
G^{(p)}_{ab;cd}(q,k) &=\left(\frac{\delta_{ac}\delta_{bd}+\delta_{ad}\delta_{bc}}{2}\right)G_1^{(p)}(q,k)+\left(\frac{\delta_{ac}+\delta_{ad}+\delta_{bc}+\delta_{bd}}{4}\right)G_2^{(p)}(q,k)+G_3^{(p)}(q,k) \ , \\
G_1^{(p)} &= \left[M_1^{(p)}\right]^{-1}   \ , \\
G_2^{(p)} &= -2 \left[ 2 M_1^{(p)}-M_2^{(p)} \right]^{-1} \otimes M_2^{(p)} \otimes \left[M^{(p)}_1 \right]^{-1} \ , \\
G_3^{(p)} &= \left[M_1^{(p)}\right]^{-1} \otimes\left\{ M_2^{(p)} \otimes \left[ 2 M_1^{(p)}-M_2^{(p)} \right]^{-1}
\otimes M_2^{(p)}-M_3^{(p)} \right\} \otimes \left[M_1^{(p)}\right]^{-1} \ .
\end{split}
\eeq
The convolution and inversion operations appearing in Eq.~\eqref{Gpqk}
are defined by
\beq\begin{split}
[A \otimes B](q,q') &= \int \frac{\de k}{(2\pi)^D} A^{(p)}(q,k) B^{(p)}(-k,q') \ , \\
[A \otimes A^{-1}](q,q')  &= (2\pi)^D\frac{\delta(q-q')+\d(q+q')}2 \ .
\end{split}\eeq

Let us now consider the four point function $G^{(f)}_{ab;cd}(r-r')$ defined in Eq.~\eqref{Gfdef}, which we can write as
\beq
G^{(f)}_{ab;cd}(r-r') = 
\int \de x \de y \, f(x)f(y) \, G_{ab;cd}\left(r+\frac{x}{2},r-\frac{x}{2};r'+\frac{y}{2},r'-\frac{y}{2}\right) \ .
\eeq
We have
\beq
\begin{split}
(2\pi)^D\delta(p+\hat p) G^{(f)}_{ab;cd}(p) &= \int \de r\de r' e^{i p r + i \hat p r'} 
\int \de x \de y \, f(x)f(y) \, G_{ab;cd}\left(r+\frac{x}{2},r-\frac{x}{2};r'+\frac{y}{2},r'-\frac{y}{2}\right) \\
 &= \int \frac{\de q \, \de k}{(2\pi)^{2D}} f(-q) f(-k) \int \de r\de r' \de x\de y \, e^{i p r + i \hat p r' + i q x + i k y} 
 \, G_{ab;cd}\left(r+\frac{x}{2},r-\frac{x}{2};r'+\frac{y}{2},r'-\frac{y}{2}\right) \\
&= \int \frac{\de q \, \de k}{(2\pi)^{2D}} f(-q) f(-k) \int_{1,2,3,4} \,
e^{ip\left(\frac{x_1+x_2}{2}\right) + \hat p \left(\frac{x_3+x_4}{2}\right)+iq(x_1-x_2)+ik(x_3-x_4)}
 \, G_{ab;cd}(1,2;3,4) \\
&=(2\pi)^D\delta(p+\hat p)  \int \frac{\de q \, \de k}{(2\pi)^{2D}} f(-q) f(-k) G^{(p)}_{ab;cd}(q,k) \ .
\end{split}
\eeq
Hence we finally obtain
\beq
G^{(f)}_{ab;cd}(p) =  \int \frac{\de q \, \de k}{(2\pi)^{2D}} f(-q) f(-k) G^{(p)}_{ab;cd}(q,k) \ .
\eeq
Using Eq.~\eqref{GdGr} we find that
\beq
G_{th}(p) =  \int \frac{\de q \, \de k}{(2\pi)^{2D}} f(-q) f(-k) \left[ G^{(p)}_{ab;ab}(q,k) - G^{(p)}_{ab;ac}(q,k) \right]
=\int \frac{\de q \, \de k}{(2\pi)^{2D}} f(-q) f(-k) \left[ \frac12 G_1^{(p)}(q,k) + \frac14 G_2^{(p)}(q,k) \right] \ .
\eeq

We can now perform some formal manipulations on the operator, dropping the indexes for convenience. Using
Eq.~\eqref{Gpqk}
\beq\begin{split}
\frac{G_1}2 + \frac{G_2}4 &=  \frac1{M_1} + \left[ \frac1{M_2 - 2 M_1} M_2 - 1 \right]\frac1{2M_1} \\
 &=  \frac1{M_1} + \left[ \frac1{M_2 - 2 M_1} M_2 - \frac1{M_2 - 2 M_1} (M_2 - 2M_1) \right]\frac1{2M_1} \\
& =  \frac1{M_1} + \left[ \frac1{M_2 - 2 M_1} 2 M_1 \right]\frac1{2M_1} \\
& =  \frac1{M_1} + \frac1{M_2 - 2 M_1}
\end{split}\eeq
so that the critical behavior is driven by the dominant divergent contribution to the kernel operator $M_1$:
\beq\label{Gthsing}
G^{sing}_{th}(p) 
=\int \frac{\de q \, \de k}{(2\pi)^{2D}} f(-q) f(-k) \left[ M_1^{(p)} \right]^{-1}(q,k) \ .
\eeq

\subsection{Spectrum of the mass matrix}
\label{sec:gradB}

Let us now study the kernel operator $M_1$ in more detail. 
We have seen in the previous sections that it develops a zero mode when approaching the transition from below. 
Moreover we know that the zero mode is translationally invariant. 
Let us now look at the spectrum of this operator. 
The crucial observation is that because of translational invariance, the mass matrix is proportional
to $\d(p+\hat p)$, so we can diagonalize 
the kernel $M^{(p)}_1(q,k)$ at fixed external momentum $p$. 
We can thus write a spectral decomposition
\beq
M_1^{(p)}(q,k)=\lambda_{0}(p^2)\psi_{0}^{(p)}(q)\psi_{0}^{(p)}(k)+\sum_{\a\geq1}\lambda_{\a}(p^2) \psi_{\a}^{(p)}(q)\psi_{\a}^{(p)}(k)
\eeq
where we put the lowest mode in evidence for future convenience and
\beq\begin{split}
&\int \frac{\de k}{(2 \p)^D}\, M_1^{(p)}(q,k)\psi_{\a}^{(p)}(-k)=\lambda_{\a}(p^2)\psi_{\a}^{(p)}(q) \ , \\
&\int \frac{\de k}{(2 \p)^D}\, \psi_{\a}^{(p)}(k)\psi_{\a'}^{(p)}(-k) = \d_{\a,\a'} \ .
\end{split}\eeq
In what follows we will suppose that there is a persistent mass gap between the zeroth eigenvalue and the first excited ones, even at the transition. This means that even if the zeroth order eigenvalue goes to zero at the transition, we are assuming that all the excited ones remain finite. Let us now consider again Eq.~\eqref{M1k0eq} 
for the zero mode
\begin{gather}
\int_{3,4}M_1(x_1,x_2,x_3,x_4)k_0(x_3-x_4)=(\mu\sqrt\epsilon +O(\epsilon))k_0(x_1-x_2) \ ;
\end{gather}
in Fourier space it is given by
\beq
\int \frac{\de k}{(2\pi)^D} M_1^{(p=0)}(q,-k)k_0(k)=(\mu\sqrt\epsilon +O(\epsilon))k_0(q) \ .
\eeq
Therefore
\beq
\psi_{0}^{(p=0)}(q)=k_0(q) \ , \ \ \ \ \lambda_{0}(p=0)=\mu\sqrt\epsilon +O(\epsilon) \ ,
\eeq
Let us now define (for $i=1,2,3$)
\beq\label{midef}
m_i = \int \frac{\de q \, \de k}{(2\pi)^{2D}}k_0(-q)M_i^{(p=0)}(q,k)k_0(-k) \ ,
\eeq
in such a way that
\beq\label{l0crit}
\lambda_0(p^2)=m_1+ \s p^2 + O(p^4) = \mu\sqrt\epsilon+\sigma p^2 +O(\epsilon, p^4) \ .
\eeq
Hence $\lambda_0(p^2)$ is small close to the transition and for small momentum $p\simeq 0$, so
that the leading term in $M_1^{-1}$ is the zero mode:
\beq\label{M1crit}
\left[M_1^{(p)}\right]^{-1}(q,k)=\frac{1}{\lambda_0(p^2)}\psi_0^{(p)}(q)\psi_0^{(p)}(k)+\sum_{\a\geq 1}
\frac{1}{\lambda_\a(p^2)}\psi_\a^{(p)}(q)\psi_\a^{(p)}(k)
\simeq \frac{1}{\lambda_0(p^2)}\psi_0^{(p)}(q)\psi_0^{(p)}(k)
\eeq
We now need to compute the coefficients $\mu$ and $\sigma$. The first one is given by the definition
\beq\label{mudef}
\mu=
\lim_{\epsilon\to 0}\frac{\de m_1}{\de\sqrt\epsilon} =
\lim_{\epsilon\to 0}\frac{\de }{\de\sqrt\epsilon}
\int \frac{\de^D q \, \de^D k}{(2\pi)^{2D}}k_0(-q)M_1^{(p=0)}(q,k)k_0(-k) \ ,
\eeq
and the other one can be evaluated using perturbation theory. In fact, let us 
develop the operator $M_1$ around $p=0$. We have that
\beq
M_1^{(p)}(q,k)=M_1^{(p=0)}(q,k)+p^2\left.\frac{\partial}{\partial p^2}M_1^{(p)}(q,k)\right|_{p=0}+O(p^4)
\eeq
and we know that $k_0(q)$ is the fundamental state of the unperturbed operator $M_1^{(p=0)}(q,k)$. Now let us treat the second term as a small perturbation because we are near $p=0$. The shift of the eigenvalue of the ground state is given by
\beq
 \lambda_0(p^2) - m_1 =
p^2\lim_{\epsilon\to 0} \int \frac{\de q \, \de k}{(2\pi)^{2D}}k_0(-q)
\left.\frac{\partial}{\partial p^2}M_1^{(p)}(q,k)\right|_{p=0}k_0(-k) \ ,
\eeq
so that
\beq\label{sigmadef}
\sigma=\lim_{\epsilon\to 0} \int \frac{\de q \, \de k}{(2\pi)^{2D}}k_0(-q)
\left.\frac{\partial}{\partial p^2}M_1^{(p)}(q,k)\right|_{p=0}k_0(-k)
 \ .
\eeq
We conclude from this analysis, and
in particular from Eq.~\eqref{l0crit} and \eqref{M1crit},
that close to the transition and for small $p$,
\beq\label{M1zeromode}
\left[M_1^{(p)}\right]^{-1}(q,k)
\simeq \frac{1}{\mu\sqrt\epsilon+\sigma p^2} k_0(q)k_0(k)
\eeq
Inserting this in Eq.~\eqref{Gthsing} we get, with the same definition of the $\star$ product as before:
\beq\label{Gsing_final}
G^{sing}_{th}(p) = \frac{(f \star k_0)^2}{\mu \sqrt{\epsilon} + \sigma p^2}  = \frac{G_0 \epsilon^{-1/2}}{1 + \xi^2 p^2} \ , \ \ \ \ \ \ \ f\star k_0 = \int \frac{\de q}{(2\pi)^D} f(-q)k_0(q) \ .
\eeq
In this way it is clear that the correlation length is given by
\beq\label{xiGth}
\xi=\sqrt{\frac{\sigma}{\mu}}\epsilon^{-1/4}
\eeq
and
\beq \label{4pointfunction_prefactor}
G_0 =  \frac{(f \star k_0)^2}{\mu}\:.
\eeq
This completes the analysis of the quadratic part of the action.

\section{Ginzburg criterion} 

In the previous section we have investigated the quadratic and cubic terms 
in an expansion of the free energy around the critical value of the order parameter.
This corresponds to the Landau expansion described in Sec.~\ref{sec:illA} and
Sec.~\ref{sec:illB}.
We now follow the discussion of Sec.~\ref{sec:illC}: we assume that the free energy
$\G$ has been obtained by truncating in some way the high temperature expansion
(we will come back on this point in Sec.~\ref{sec:HNC}), and we use
this mean field free energy
as a bare action to perform a loop computation and obtain a Ginzburg criterion.

The quadratic part of the free energy provides the Gaussian part of
the bare action. 
What can be expected from the discussion of Sec.~\ref{sec:illD} 
is that the Gaussian approximation is valid (at least qualitatively) everywhere in the 
$(T,\r)$ plane if the dimension is greater than the upper critical dimensions;
otherwise the Gaussian approximation is valid only up to a certain distance 
from the critical line in the $(T,\r)$ plane. 
As discussed in Sec.~\ref{sec:illD}, this statement 
can be made more precise by performing a one loop expansion 
for the correlation function of the order parameter and then look at which 
temperature or density the first loop correction is of the same order of the 
Gaussian approximation for that correlation.
The Gaussian bare propagator is $G^{(p)}_{ab;cd}(q,k)$
given in Eq.~\eqref{Gpqk}.
To compute the loop corrections to this propagator, one should start a very difficult 
computation because the diagrammatic arising from the replica field theory is complicated 
by the presence of the replica indices as well as the ``internal'' wavevectors $q,k$. 

\subsection{Projection on the zero mode}

The calculation can be greatly simplified if we focus only on the critical behavior of the correlation functions.
In particular, in Sec.~\ref{sec:gradB} we have shown that the critical part of the propagator $G_{ab;cd}^{(p)}(q,k)$ is entirely 
dominated by the presence of the zero mode of the kernel operator $M_1^{(p)}(q,k)$. For example, according to Eqs.~\eqref{Gthsing}
and \eqref{M1zeromode},
the singular part of the 
thermal correlation function $G_{th}(p)$ is given by the inverse of the zero mode of this operator. 
Clearly, the full thermal correlation function contains also the non critical contribution 
coming from the excited states but here we are not interested in this contributions. 
Instead, we would like to focus on the critical part of the observables. To do this we can suppose 
that the eigenvalues corresponding to excited states are set to infinity so that we have no 
fluctuations along the directions orthogonal to the zero mode. 
This amounts to consider an effective low-energy theory where only the fluctuations along the zero mode are allowed, which means 
that we choose
\beq\label{Deltarzeromode}
\Delta\r_{ab}^{(2)}(p,q)=\phi_{ab}(p)k_0(q) \ .
\eeq 
By doing this we obtain a simplified theory that gives us only the critical part of the correlation we want to compute. 
The resulting effective theory is obtained as follows. The quadratic part is obtained by inserting Eq.~\eqref{Deltarzeromode} in 
Eq.~\eqref{D2Ggrad}. The relevant cubic terms according to~\cite{TDP02,FPRR11} can be obtained by inserting Eq.~\eqref{Deltarzeromode} 
in the cubic part of the expansion and using similar considerations as in Sec.~\ref{sec:LandauB}. The result is
\beq
\begin{split}\label{EFT}
\Gamma[ \phi_{ab}]&=\frac{1}{2}\int \frac{\de p}{(2\pi)^D} \;\left(
\sum_{a\neq b} (\mu\sqrt{\epsilon}+\sigma p^2) |\phi_{ab}(p)|^2 +m_2 \sum_a\left|\sum_b \phi_{ab}(p) \right|^2 +m_3 \left|\sum_{a\ne b} \phi_{ab}(p) \right|^2 \right)\\
&+\frac{w_1}{6} \sum_{a\neq b\neq c\neq a} \int\frac{\de p\de p'}{(2\pi)^{2D}} \phi_{ab}(p)\phi_{bc}(p')\phi_{ca}(-p-p') +\frac{w_2}{6} \sum_{a\neq b} 
\int\frac{\de p\de p'}{(2\pi)^{2D}}\phi_{ab}(p)\phi_{ab}(p')\phi_{ab}(-p-p') \ ,
\end{split}
\eeq 
and the coefficients that enters in the above formula are exactly the ones that were computed in the previous sections:
see Eqs.~\eqref{midef}, \eqref{mudef} and \eqref{sigmadef} for the quadratic part and Eq.~\eqref{w1_w2} (without the factor $f\star k_0$ in the denominator) for the
cubic part. This result is at the basis of our next computation. 
Let us note that by projecting on the zero mode, the momentum structure of the theory has been simplified drastically.

\subsection{$\f^3$ theory in random field}

To get the result at the dynamical transition, one should take the $m\to 1$ limit. 
In \cite{FPRR11} it has been shown that the perturbative expansion of the replica field theory of the form (\ref{EFT}) 
can be mapped to the study of a scalar field which satisfies a particular cubic stochastic field equation. 
This mapping has been done using a transformation for the fields which is quite close to the Cardy's treatment 
of the branched polymer problem~\cite{Ca83}. 
However using the field theory techniques developed by Parisi and Sourlas~\cite{PS82}, 
it can be shown straightforwardly that the stochastic equation describes also the leading critical behavior of a theory 
for a scalar field in a cubic potential and interacting with a random Gaussian magnetic field. 
The action of this spinodal field theory 
in a random field --which we now use as a bare action following the discussion of Sec.~\ref{sec:illC}-- is given by
\begin{gather}\label{RFaction}
S(\f)=\frac{1}{2}\int \de x \,\f(x)(-\nabla^2+m_0^2 )\f(x)+\frac{g}{6}\int\de x\f^3(x)+\int\de x ( h_0(x) + \d h(g,\D) )\f(x)
\end{gather}
where $h_0(x)$ is a gaussian random field with variance given by
\beq
\overline{h_0(x)h_0(y)}=\Delta\delta(x-y) \ ,
\eeq
and the coupling and masses are given in terms of the coupling and masses that appear in the replica field theory (\ref{EFT}) by 
\beq \label{mass^2}
\begin{split}
m_0^2&= \mu\sqrt\epsilon / \sigma \ , \\
g&= (w_1 - w_2)/\sigma^{3/2} \ , \\
\Delta&=-(m_2+m_3) / \sigma \ .
\end{split}
\eeq
In the action (\ref{RFaction}) we added a counterterm $\d h$, that will be used to avoid that one loop corrections shift 
the position of the critical point. The counterterm $\d h$ can be seen also as a redefinition of the mean value for the 
random field $h_0$. The bare propagator of the theory is, as usual, given by $G_0(p) = (p^2+m_0^2)^{-1}$. 
To compute loop corrections, it is quite useful to write down the generating functional for the connected diagrams
\beq
W[J]=\ln Z[J] \ , 
\eeq
where the external current $J(x)$ is given by
\beq
J(x)=h_0(x)+\delta h(g,\Delta) \ .
\eeq
Moreover, let us introduce the diagrammatic notation that will be convenient later
\beq
\begin{picture}(30,15)(-15,-2)
\SetColor{Black}
\SetWidth{1}
\CCirc(0,0){2}{Black}{White}
\end{picture}
=
\begin{picture}(30,15)(-15,-2)
\SetColor{Black}
\SetWidth{1}
\CCirc(0,0){2}{Black}{Black}
\end{picture}
+
\begin{picture}(30,15)(-15,-2)
\SetColor{Black}
\SetWidth{1}
\CCirc(0,0){2}{Red}{Red}
\end{picture}
\eeq
where
\beq
J(x)=\begin{picture}(30,15)(-15,-2)
\SetColor{Black}
\SetWidth{1}
\CCirc(0,0){2}{Black}{White}
\end{picture}
\ \ \ \ \ \ \ 
h_0(x)=\begin{picture}(30,15)(-15,-2)
\SetColor{Black}
\SetWidth{1}
\CCirc(0,0){2}{Black}{Black}
\end{picture}
\ \ \ \ \ \ \ 
\d h(g,\Delta)=\begin{picture}(30,15)(-15,-2)
\SetColor{Black}
\SetWidth{1}
\CCirc(0,0){2}{Red}{Red}
\end{picture}
\eeq
The expansion for $W[J]$ is given (up to second order in $g$) by the following diagrams
\begin{gather}
W[J]=
\begin{picture}(30,60)(-15,-2)
\SetColor{Black}
\SetWidth{1}
\Line(0,0)(0,20)
\CArc(0,30)(10,0,360)
\CCirc(0,0){2}{Black}{White}
\end{picture}
+\begin{picture}(45,60)(-22.5,-2)
\SetColor{Black}
\SetWidth{1}
\Line(-15,0)(15,0)
\CCirc(-15,0){2}{Black}{White}
\CCirc(15,0){2}{Black}{White}
\end{picture}
+\begin{picture}(45,60)(-22.5,-2)
\SetColor{Black}
\SetWidth{1}
\Line(0,0)(0,20)
\Line(-15,0)(15,0)
\CCirc(-15,0){2}{Black}{White}
\CCirc(15,0){2}{Black}{White}
\CCirc(0,20){2}{Black}{White}
\end{picture}
+\ \ 
\begin{picture}(60,60)(-30,-2)
\SetColor{Black}
\SetWidth{1}
\Line(-30,0)(30,0)
\CCirc(-30,0){2}{Black}{White}
\CCirc(30,0){2}{Black}{White}
\CArc(0,0)(18,0,180)
\end{picture}
\ \ +\ \ 
\begin{picture}(30,60)(-15,-2)
\SetColor{Black}
\SetWidth{1}
\Line(-15,0)(15,0)
\Line(0,0)(0,20)
\CArc(0,30)(10,0,360)
\CCirc(-15,0){2}{Black}{White}
\CCirc(15,0){2}{Black}{White}
\end{picture}
\ \ +\ \ 
\begin{picture}(60,60)(-30,-2)
\SetColor{Black}
\SetWidth{1}
\Line(-30,0)(30,0)
\CCirc(-30,0){2}{Black}{White}
\CCirc(30,0){2}{Black}{White}
\Line(0,0)(0,20)
\Line(0,20)(-15,35)
\Line(0,20)(15,35)
\CCirc(-15,35){2}{Black}{White}
\CCirc(15,35){2}{Black}{White}
\end{picture}
\ \ +\ldots
\end{gather}
Let's look first to the corrections to the average of $\la \f(x) \ra$. 
By taking the average over the random field and by marking with a blue dot a doubled propagator, these corrections are given by
\beq
\overline{\la \f(x) \ra} = 
\begin{picture}(30,60)(-15,-2)
\SetColor{Black}
\SetWidth{1}
\Line(0,0)(0,20)
\CCirc(0,20){2}{Red}{Red}
\end{picture}
+ 
\begin{picture}(30,60)(-15,-2)
\SetColor{Black}
\SetWidth{1}
\Line(0,0)(0,20)
\CArc(0,30)(10,0,360)
\CCirc(0,40){2}{Blue}{Blue}
\end{picture}
+
\begin{picture}(30,60)(-15,-2)
\SetColor{Black}
\SetWidth{1}
\Line(0,0)(0,20)
\CArc(0,30)(10,0,360)
\end{picture}
\eeq
where we have made the hypothesis (to be checked in a self consistent way) that the counterterm $\delta h$ 
is proportional to $g$ so that we have retained only the diagrams up to second order in $g$. 
We want that the critical point is not shifted so we want $\overline{\la \f(x) \ra}=0$. 
Therefore at this order in perturbation theory we must have
\beq
\begin{picture}(30,60)(-15,-2)
\SetColor{Black}
\SetWidth{1}
\Line(0,0)(0,20)
\CCirc(0,20){2}{Red}{Red}
\end{picture}
+ 
\begin{picture}(30,60)(-15,-2)
\SetColor{Black}
\SetWidth{1}
\Line(0,0)(0,20)
\CArc(0,30)(10,0,360)
\CCirc(0,40){2}{Blue}{Blue}
\end{picture}
+
\begin{picture}(30,60)(-15,-2)
\SetColor{Black}
\SetWidth{1}
\Line(0,0)(0,20)
\CArc(0,30)(10,0,360)
\end{picture}=0 \label{cancellation}
\eeq
from which we see that $\delta h\propto g$. Now we look to the correction to the propagator. We have
\begin{align*}
\overline{\la \f(x) \f(y) \ra} &= G_0(x-y) + 
\begin{picture}(50,40)(-25,-2)
\SetColor{Black}
\SetWidth{1}
\Line(-25,0)(25,0)
\Line(0,0)(0,20)
\CCirc(0,20){2}{Red}{Red}
\end{picture} 
+ \begin{picture}(30,60)(-15,-2)
\SetColor{Black}
\SetWidth{1}
\Line(-15,0)(15,0)
\Line(0,0)(0,20)
\CArc(0,30)(10,0,360)
\end{picture}
+ \begin{picture}(30,60)(-15,-2)
\SetColor{Black}
\SetWidth{1}
\Line(-15,0)(15,0)
\Line(0,0)(0,20)
\CArc(0,30)(10,0,360)
\CCirc(0,40){2}{Blue}{Blue}
\end{picture}
+
\begin{picture}(50,40)(-25,-2)
\SetColor{Black}
\SetWidth{1}
\Line(-25,0)(25,0)
\CArc(0,0)(15,0,180)
\CCirc(0,15){2}{Blue}{Blue}
\end{picture}+	
\begin{picture}(50,40)(-25,-2)
\SetColor{Black}
\SetWidth{1}
\Line(-25,0)(25,0)
\CArc(0,0)(15,0,180)
\end{picture}+\ldots \\
&= G_0(x-y) + 
\begin{picture}(50,40)(-25,-2)
\SetColor{Black}
\SetWidth{1}
\Line(-25,0)(25,0)
\CArc(0,0)(15,0,180)
\CCirc(0,15){2}{Blue}{Blue}
\end{picture} +
\begin{picture}(50,40)(-25,-2)
\SetColor{Black}
\SetWidth{1}
\Line(-25,0)(25,0)
\CArc(0,0)(15,0,180)
\end{picture}+\ldots
\end{align*}
where we have used the relation (\ref{cancellation}). Because we are interested in the most (infrared) divergent diagrams we can neglect the second one. However, note that the second diagram is relevant in the ultraviolet regime because it diverges sooner as the number of dimensions is increased (actually it is convergent in $D<4$). Now, let us neglect the second diagram and compute the dotted diagram. We have at zero momentum
\begin{gather}
G(p=0) = G_0(p=0) + G_0(p=0)^2 \frac{\D g^2}{2}\int^\L \frac{\de q}{(2\pi)^D} G_0(q)^3 \ .
\end{gather}
From now on we follow closely the procedure of Sec.~\ref{sec:illD}.
We invert the above equation to obtain
\begin{gather}\label{mRm03}
m^2_R = G^{-1}(p=0) = m_0^2 -  \frac{\D g^2}{2}\int^\L\frac{\de q}{(2\pi)^D} \frac1{(q^2 + m_0^2)^3}
= m_0^2 -  \frac{\D g^2}{2}\int^\L\frac{\de q}{(2\pi)^D} \frac1{(q^2 + m_R^2)^3} \ .
\end{gather}
As in the $\f^4$ theory, the mass correction is UV divergent for $d\geq 6$ and 
leads to a non-universal shift of the critical point,
which corresponds to
\beq
m_0^2 =  \frac{\D g^2}{2}\int^\L\frac{\de q}{(2\pi)^D} \frac1{q^6} \ .
\eeq
The distance from the critical point is
\beq
t = m_0^2 - \frac{\D g^2}{2}\int^\L\frac{\de q}{(2\pi)^D} \frac1{q^6} \ ,
\eeq
and Eq.~\eqref{mRm03} becomes
\beq
m^2_R = t -  \frac{\D g^2}{2}\int^\L\frac{\de q}{(2\pi)^D} \left(\frac1{(q^2 + m_R^2)^3} - \frac1{q^6}\right) \ .
\eeq
As in Sec.~\ref{sec:illD} we have to impose the condition that the one loop correction is small
in such a way that the mean field result $m^2_R =t$ is not altered.
A simpler and completely equivalent (apart from a numerical prefactor of order one) 
condition is obtained by considering
\beq
\frac{d t}{d m^2_R} = 1  - 3   \frac{\D g^2}{2}\int^\L\frac{\de q}{(2\pi)^D}  \frac1{(q^2 + m_R^2)^4} \ ,
\eeq
and of course we want the second term to be much smaller than one.
Written in terms of physical observables ($m_R^2$ and $t$), 
the one loop correction is UV convergent and IR divergent
when $D<8$ and the renormalized theory exists in this case, 
leading to a universal form of the Ginzburg criterion,
which clearly tells us that the upper critical dimension is 8. 
Sending the cutoff to infinity, using the standard Schwinger representation for the propagator~\cite{ParisiBook}
\begin{gather*}
\frac{1}{p^2+m^2}=\int_0^\infty \de \alpha \,e^{-\alpha(p^2+m^2)}
\end{gather*}
and taking derivatives with respect to $m^2$, we obtain
\begin{gather}
\frac{d t}{d m^2_R} = 1  - \frac{\Delta g^2}{4(4\pi)^{D/2}}\Gamma\left(4-\frac{D}{2}\right) m_R^{D-8}
\end{gather} 
and the Ginzburg Criterion amounts to the following condition:
\begin{gather}
1\gg \frac{g^2\Delta}{4(4\pi)^{D/2}} \Gamma\left(4-\frac{D}{2}\right) m_R^{D-8} \ .
\end{gather}
On the other hand for $D\geq 8$, as before, the correction is UV divergent (and IR convergent) and therefore the precise form of the Ginzburg criterion depends on the regularization. 
Again in the mean field region the correlation length is $\xi = 1/m_R$, see Eq.~\eqref{xiGth}, hence we can write 
\beq \label{ginzburg}
1\gg \textrm{Gi}\, \xi^{8-D} \ ,
\hskip50pt
\textrm{Gi}=\frac{g^2\Delta}{4(4\pi)^{D/2}}\Gamma\left(4-\frac{D}{2}\right) \ .
\eeq

\section{Explicit calculations in the Hypernetted Chain (HNC) approximation}
\label{sec:HNC}

The theory developed in the previous sections allows to compute several observables related to the critical behavior of the
system at the dynamical transition. The starting point of the theory, that is needed
in order to perform concrete microscopic computations, is an explicit ``mean-field'' expression for the
free energy $\G[\r,\r^{(2)}]$. As discussed in Sec.~\ref{sec:illC}, there are several possibilities
and many of them have been used in liquid theory~\cite{Hansen}: most of them can be seen as resummation
of the high temperature or low density expansion. 
Once a mean-field approximation to the free energy has been chosen, the observables can be computed. Before proceeding,
let us summarize the main results of the previous sections.
\begin{itemize}
\item First of all one has to identify the dynamical transition point $T_d$ or $\r_d$, at which the off diagonal part $\wt g$ of the physical
correlation defined in Eq.~\eqref{SP} jumps from the trivial value $\wt g = 1$ to a non-trivial value.
\item The next step is to identify the zero mode $k_0$ of the mass operator, as discussed in Sec.~\ref{sec:LandauC}. 
This can be done by a direct diagonalization of the mass operator, defined in Eqs.~\eqref{mass_def2} and \eqref{mass}. Recall
that according to the analysis of Sec.~\ref{sec:LandauC} the zero mode is determined only by the $M_1$ term of the mass operator,
see Eq.~\eqref{M1k0eq}. However, a simpler route, which we will use in the following, is to determine the evolution of $\wt g$ close to the
critical point and extract the zero mode from Eq.~\eqref{sqrtSP}.
\item At this point we know the zero mode, the free energy and therefore its second ($M$) and third ($L$) derivatives, see Eq.~\eqref{mass_and_cubic_interactions}.
The first observable we can compute is the exponent parameter $\l$
of Mode-Coupling Theory. We have $\l = w_2/w_1$ where the cubic coefficients $w_1$ and $w_2$ are given by Eq.~\eqref{w1_w2}.
Here the choice of the smoothing function $f$ that is used to define a proper overlap (see the discussion in Sec.~\ref{sec:defA}) is irrelevant, because
the ratio of $w_1$ and $w_2$ is independent of $f$.
\item Next we can look at the gradient expansion for the fluctuations along the zero mode. We compute the masses $m_2$ and $m_3$ from
Eq.~\eqref{midef}. The critical part is once again related to the zero mode of $M_1$, as expressed in Eqs.~\eqref{l0crit} and \eqref{M1crit}. We need the two coefficients
$\mu$ and $\sigma$, given respectively by Eqs.~\eqref{mudef} and \eqref{sigmadef}. From these coefficients we obtain the singular part of the
four-point correlations, Eq.~\eqref{Gsing_final}, with the dynamical correlation given in Eq.~\eqref{xiGth} and
the $f$-dependent prefactor given in Eq.~\eqref{4pointfunction_prefactor}.
\item At this point we have all the couplings that enter in the effective action for the fluctuations along the zero mode, Eq.~\eqref{EFT}. 
Remember that here we choose $f=k_0$, hence the $f$ dependent denominators in $w_1$ and $w_2$, Eq.~\eqref{w1_w2}, are omitted.
Mapping this theory on a $\f^3$ theory with the change of couplings in Eq.~\eqref{mass^2}, we obtain the Ginzburg criterion expressed by
Eq.~\eqref{ginzburg}. 
\end{itemize}

In order to present a concrete implementation of this program, in this section we present explicit calculations using
a popular approximation for the free energy, the so-called
Hypernetted Chain (HNC) approximation. It amounts to simply 
neglect the contribution of 2PI diagrams from Eq.~(\ref{GammaGen}).
This approximation is known to give good qualitative results
for the structure functions around the dynamical transition~\cite{MP96,CFP98,PZ10}. 
However, the quantitative agreement with numerical results is quite poor, in particular for the so-called non-ergodicity
factor, which is related to the Fourier transform of $\wt g$. We will see that in fact the results for $\l$ are not very good.
Yet we believe that the order of magnitude for the Ginzburg number, which is the most interesting result of the present analysis,
is correct.

Better approximations can be obtained by performing a ``small
cage expansion''~\cite{MP09,PZ10} or by expanding systematically around the HNC equations~\cite{JZ12}.
Other approximation schemes have been explored~\cite{Sz10,Ri12} but they are not practical for the present
purposes. A preliminary calculation of $\l$ using the small cage expansion~\cite{KPUZ13} gives very good agreement
with the numerical result. However, the aim of this section is only to show that concrete calculations can be done easily:
we leave a detailed comparison with numerical results for future work.

\subsection{Derivatives of the free energy in HNC}

Because we need the expansion of the free energy up to the third order we start by computing its derivatives. 
To simplify the notations, here we will first perform the calculation without taking into account the symmetries
of the replica and spatial indices, and we will symmetrize the result at the end. We also use the shorthand
notation $\d_{ab}(1,2) = \d_{ab} \d(x_1-x_2)$.
 
The first derivative is
\beq
\frac{\d \G[\r,\r^{(2)}]}{\d \r^{(2)}_{ab}(1,2)} = \frac12 \log\left[ \frac{\r^{(2)}_{ab}(1,2)}{\r_a(1) \r_b(2)} \right]
+ \frac12 \left[ c_{ba}(2,1) - h_{ba}(2,1) \right] \ ,
\eeq
which can be easily proved from Eq.~\eqref{Gammaidring} by 
representing $\G_{\rm ring}$ as a sum of diagrams~\cite{Hansen}.
Using the same diagrammatic representation we find
\beq
\frac{\d c_{ab}(1,2)}{\d h_{cd}(3,4)} = \big[ \d_{ac}(1,3) - \r_c(3) c_{ac}(1,3) \big]  \big[ \d_{db}(4,2) - \r_d(4) c_{db}(4,2) \big] 
=\r_c(3) \r_d(4)  \G^{(2)}_{ac}(1,3) \G^{(2)}_{db}(4,2) 
\eeq
having defined
\beq
\G^{(2)}_{ab}(1,2) = \frac{1}{\r_a(1)} \d_{ab}(1,2) - c_{ab}(1,2) \ . \label{gamma2}
\eeq
From this we can compute the second derivative of the HNC free energy, namely its mass matrix:
\beq\label{gamma02_HNC}
\begin{split}
\frac{\d^2 \G[\r,\r^{(2)}]}{\d \r^{(2)}_{ab}(1,2)\d \r^{(2)}_{cd}(3,4)}  = & \frac12   \left[ \frac{ \d_{ac}(1,3) \d_{db}(4,2) }{\r^{(2)}_{ab}(1,2)}  -  \frac{ \d_{ad}(1,4) \d_{bc}(2,3) }{\r_{a}(1) \r_b(2)} \right] \\
&+ \frac12 \frac{1}{\r_c(3) \r_d(4)} \big[ \d_{bc}(2,3) - \r_c(3) c_{bc}(2,3) \big]  \big[ \d_{da}(4,1) - \r_d(4) c_{da}(4,1) \big] \\
= & \frac12   \left[ \frac{ \d_{ac}(1,3) \d_{db}(4,2) }{\r^{(2)}_{ab}(1,2)}  -  \frac{ \d_{ad}(1,4) \d_{bc}(2,3)}{\r_{a}(1) \r_b(2)} \right]  + \frac12 \G^{(2)}_{bc}(2,3)\G^{(2)}_{da}(4,1)\:.
\end{split}
\eeq
Moreover the cubic terms are given by
\beq\begin{split}
\frac{\d^3 \G[\r,\r^{(2)}]}{\d \r^{(2)}_{ab}(1,2)\d \r^{(2)}_{cd}(3,4)\d \r^{(2)}_{ef}(5,6)}  = & -\frac12  \d_{ac}(1,3) \d_{db}(4,2) \d_{ae}(1,5) \d_{fb}(6,2)  \frac{1 }{[\r^{(2)}_{ab}(1,2)]^2}   \\
&- \frac12 \frac{1}{ \r_c(3)\r_e(5) \r_f(6)} \big[ \d_{bc}(2,3) - \r_c(3) c_{bc}(2,3) \big]  \frac{\d c_{da}(4,1)}{\d h_{ef}(5,6)} \\
&- \frac12 \frac{1}{ \r_d(4)\r_e(5) \r_f(6)} \frac{\d c_{bc}(2,3)}{\d h_{ef}(5,6)}  \big[ \d_{da}(4,1) - \r_d(4) c_{da}(4,1) \big]  \\
= & -\frac12  \d_{ac}(1,3) \d_{db}(4,2) \d_{ae}(1,5) \d_{fb}(6,2)  \frac{1 }{[\r^{(2)}_{ab}(1,2)]^2}   \\
&- \frac12 \left[  \G^{(2)}_{bc}(2,3)  \G^{(2)}_{de}(4,5)  \G^{(2)}_{fa}(6,1) +  \G^{(2)}_{be}(2,5)  \G^{(2)}_{cf}(3,6)  \G^{(2)}_{da}(4,1) \right]
\end{split} \label{gamma03_HNC}
\eeq

Using these results, following Sec.~\ref{sec:LandauA}, we expand the free energy around the 1RSB 
reference solution, Eq.~\eqref{SP}. We have $\D \rd_{ab}(1,2) =  \rd_{ab}(1,2) - \overline \rd_{ab}(1,2)$
and $\D\G = \G[\r,\rd] - \G[\r,\overline\rd]$.
The derivatives of $\G$ and in particular the functions $\G^{(2)}$ 
must be evaluated on the 1RSB solution Eq.~(\ref{SP}), 
hence
\beq
\G^{(2)}_{ab}(1,2) = \frac1\r \d_{ab}(1,2) - \d_{ab} [c(1,2) -\wt c(1,2)]  - \wt c(1,2) \ ,
\eeq
and we have introduced for the matrix $c_{ab}(1,2)$ the same notation as in Eq.~(\ref{SP}), 
where $c(1,2)$ and $\tilde c(1,2)$ are respectively its diagonal and off-diagonal part.
We obtain $\D\G =  \D^2 \G + \D^3 \G$ with
\beq\label{HNCD2D3}
\begin{split}
\D^2 \G & = \frac1{4 \r^2} \sum_{a\neq b} \int_{1,2}  \left[ \frac{1 }{\wt g(1,2)}  -  1 \right] [ \D \r_{ab}(1,2) ]^2
+ \frac14 \sum_{a\neq b, c\neq d} \int_{1,2,3,4} \D\r_{ab}(1,2)  \G^{(2)}_{bc}(2,3) \D\r_{cd}(3,4)    \G^{(2)}_{da}(4,1) \ , \\
 \D^3 \G & =
- \frac1{12 \r^4} \sum_{a\neq b} \int_{1,2} \frac{1 }{[\wt g(1,2)]^2} [\D \r_{ab}(1,2) ]^3 \\
& -\frac16  \sum_{a\neq b, c\neq d, e\neq f} \int_{1,2,3,4,5,6} \D\r_{ab}(1,2)  \G^{(2)}_{bc}(2,3) \D\r_{cd}(3,4)    \G^{(2)}_{de}(4,5) \D\r_{ef}(5,6)    \G^{(2)}_{fa}(6,1) \ .
\end{split}\eeq
Remember that the above Eq.~\eqref{gamma02_HNC} and \eqref{gamma03_HNC} are not symmetrized, but obviously
when they are inserted in the free energy to compute $\D^2\G$ and $\D^3\G$ they are contracted with
symmetric functions so the result is correct. However, in the following section we will have to symmetrize
them explicitly in order to insert them in the expressions for the coefficients of the action, where
the symmetry properties have been used explicitly.

\subsection{The HNC mass matrix}

The mass matrix, due to the replica symmetry of Eq.~\eqref{SP}, can be put in the form (\ref{mass})
by a proper symmetrization of indeces in Eq.~\eqref{gamma02_HNC}.
The parameters entering in that expression are given by
\beq\label{HNCM1M2M3}
\begin{split}
M_1(1,2;3,4)&=\frac{1}{2\r^2}\frac{\delta(1,3)\delta(2,4)}{\tilde g(1,2)}-\frac{1}{2\r}\left[\delta(1,4)\Delta c(2,3)+\delta(2,3)\Delta c(1,4)\right]+\frac{1}{2}\Delta c(1,4)\Delta c(2,3) \\
M_2(1,2;3,4)&=-\frac{1}{2}\tilde c(1,4)\left(\frac{1}{\r}\delta(2,3)-\Delta c(2,3)\right)-\frac 12 \tilde c(2,3)\left(\frac{1}{\r}\delta(1,4) - \Delta c(1,4)\right) \\
M_3(1,2;3,4)&=\frac 12 \tilde c(1,4)\tilde c(2,3)
\end{split}\eeq
where we have introduced the notation
\beq
\Delta c(1,2)= c(1,2)-\tilde c (1,2) \ .
\eeq
Note that we did not symmetrize Eq.~\eqref{gamma02_HNC} over spatial indeces (i.e. over exchanges
$1\leftrightarrow 2$ and $3\leftrightarrow 4$). This has not to be done explicitly because
in any case we are going to apply these operators to symmetric functions only.

\subsection{Expression of $\l$ in HNC}

We now consider the cubic term. We want to plug Eq.~\eqref{gamma03_HNC} into
Eqs.~(\ref{w1_w2}) to obtain $w_1$ and $w_2$.
Here we have again to symmetrize Eq.~\eqref{gamma03_HNC} with respect to
the exchanges $a \leftrightarrow b$, $c \leftrightarrow d$, $e \leftrightarrow f$,
$ab \leftrightarrow cd$, $ab \leftrightarrow ef$, $cd \leftrightarrow ef$ because these
have been used explicitly to derive Eqs.~(\ref{w1_w2}). Indeed, they have been used in
Eqs.~\eqref{DeDominicis} and \eqref{DeDominicis2}.
Note that, once again, what is important is to symmetrize the replica indices, because
the spatial indices are going to be contracted with a symmetric function in Eqs.~(\ref{w1_w2})
so there is no need to symmetrize them explicitly.

Let us denote the two terms in Eq.~\eqref{gamma03_HNC} with names that highlight the
different topologies of their replica indices connections:
\beq\begin{split}
L^{(1)}_{ace,bdf}(1,2;3,4;5,6) &=  
\d_{ac}(1,3) \d_{db}(4,2) \d_{ae}(1,5) \d_{fb}(6,2)  \frac{1 }{[\r^{(2)}_{ab}(1,2)]^2} \ , \\
L^{(2)}_{bc,de,fa}(1,2;3,4;5,6) &=  
 \G^{(2)}_{bc}(2,3)  \G^{(2)}_{de}(4,5)  \G^{(2)}_{fa}(6,1) \ . \\
\end{split}\eeq
Then the symmetrized derivative is (omitting the irrelevant spatial indices)
\beq\begin{split}
L_{ab,cd,ef} =& -\frac18 \left( L^{(1)}_{ace,bdf} + L^{(1)}_{acf,bde}+ L^{(1)}_{adf,bce}+ L^{(1)}_{ade,bcf} \right) \\
&- \frac18  \left(
L^{(2)}_{ac,bf,de} +  L^{(2)}_{ac,be,df}
 +  L^{(2)}_{ad,be,cf} +  L^{(2)}_{ad,bf,ce}
 +  L^{(2)}_{ae,bc,df} +  L^{(2)}_{ae,bd,cf}
 +  L^{(2)}_{af,bc,de} +  L^{(2)}_{af,bd,ce}
\right)
\end{split}\eeq
From this the eight independent elements of $L_{ab,cd,ef}$ that enter in Eq.~\eqref{w1_w2} are easily
computed and we get the following two expressions for $w_1$ and $w_2$
\beq\label{w1_w2_HNC}
\begin{split}
w_2&=-\frac{1}{16}\frac{1}{(f\star k_0)^3 \, V}\int_{1,2;3,4;5,6}k_0(1,2)k_0(3,4)k_0(5,6)\delta(1,3)\delta(2,4)\delta(1,5)\delta(2,6)\left(\frac{1}{\rho^4\tilde g^2(1,2)}\right) \ , \\
w_1&=-\frac{1}8\frac{1}{(f\star k_0)^3 \, V}\int_{1,2,3,4,5,6}k_0(1,2)k_0(3,4)k_0(5,6)\left[\Gamma(2,3)\Gamma(4,5) (\Gamma(6,1)-3\tilde \Gamma(6,1))+
(3\Gamma(2,3)-\tilde\Gamma(2,3))\tilde\Gamma(4,5)\tilde\Gamma(6,1)\right] \ ,
\end{split}\eeq
where we have denoted $\Gamma(1,2)$ and $\tilde \Gamma(1,2)$ respectively the diagonal and off diagonal part of the matrix (\ref{gamma2}).
It follows that the exponent parameter is given by
\beq \label{lambda_HNC}
\lambda=\frac12 \frac{\int \de r \frac{k_0(r)^3}{\r^4 \tilde{g}(r)^2}}{\int \frac{\de q}{(2 \pi)^D} 
k_0(q)^3\left[\Gamma(q)-\tilde\Gamma(q)\right]^3}=
\frac12 \frac{\frac{1}{\r^4}\int \de r \frac{k_0^3(r)}{\tilde g^2(r)}}{\frac{1}{\r^3}\int \frac{\de q}{(2\pi)^D}k_0^3(q)\left[1-\r \Delta c(q)\right]^3}
\eeq

\subsection{Computation of $\mu$ and $\sigma$ in HNC}

We will now compute the coefficients of the mass matrix ($\mu, \s, m_2, m_3$) in the HNC approximation. 
The contraction of the operator $M_1^{(p)}$ with the zero mode gives
\beq\begin{split}
m_1(p) &=\int \frac{\de q\de k}{(2\pi)^{2D}}k_0(q)M_1^{(p)}(q,k)k_0(k) \\
&=\frac{1}{2\rho^2}\int \de r \frac{k_0^2(r)}{\tilde g(r)}-\frac{1}{2\rho}\int \frac{\de q}{(2\pi)^D}k^2_0(q)\left[\Delta c\left(\frac{p}{2}+q\right)+\Delta c\left(\frac{p}{2}-q\right)\right] \\
&+\frac{1}{2}\int \frac{\de q}{(2\pi)^D}k^2_0(q)\Delta c\left(\frac{p}{2}+q\right)\Delta c\left(\frac{p}{2}-q\right) \ ,
\end{split}\eeq
so the the $\mu$ coefficient defined in Eq.~\eqref{mudef} is given by
\beq
\mu=\lim_{\epsilon\to 0}\frac{\de m_1(p=0)}{\de \sqrt \epsilon} \ .
\eeq
In order to compute this derivative we recall that from Eq.~\eqref{sqrtSP}
\beq
\kappa k_0(r)  =\lim_{\epsilon\to 0}\sqrt\epsilon\frac{\de \tilde g(r)}{\de \epsilon}\implies \tilde g(r,\epsilon)=\tilde g(r,0)+2\sqrt\epsilon \kappa k_0(r)+O(\epsilon) \ .
\eeq
Then we can use the replicated Ornstein-Zernike relation~\cite{PZ10} to obtain
\beq
g(q)-\tilde g(q)=\frac{\Delta c(q)}{1-\rho\Delta c(q)} \ ,
\eeq
from which
\beq\begin{split}
\tilde c(q,\epsilon) &=\tilde c(q,0)+\sqrt\epsilon c_0(q)+O(\epsilon) \\
c_0(q)&=\lim_{\epsilon\to 0}\frac{\de \tilde c(q)}{\de \sqrt{\epsilon}}=2\kappa k_0(q)\left[1-\rho\Delta c(q)\right]^2 \ .
\end{split}\eeq
Using these expressions we arrive to the final form for the coefficient $\mu$:
\begin{gather}
\mu=-\frac{\kappa}{\rho^2} \int \de r \frac{k_0^3(r)}{\tilde g^2(r)}+\frac{2\kappa}{\rho} \int \frac{\de q}{(2\pi)^D}k_0^3(q)\left[1-\rho\Delta c(q)\right]^3 \ .
\end{gather}
In an analogous way we can obtain the expression for the coefficient $\sigma$
\beq
\sigma=\lim_{\epsilon\to 0}\left.\frac{\de m_1(p)}{\de p^2}\right|_{p=0}\:.
\eeq
To compute this expression let us use the following relation
\begin{gather}
f\left(\left|\frac{p}{2}+q \right|\right)\simeq f(q)+\frac{1}{2}\frac{f'(q)}{q}(q\cdot p)+\frac12\left[\frac14 \frac{f''(q)}{q^2}(q\cdot p)^2+\frac14 \frac{f'(q)}{q}p^2-\frac14 \frac{f'(q)}{q^3}(q\cdot p)^2\right]
\end{gather}
so that the final expression for $\sigma$ is given by
\begin{gather}
\sigma=\frac{1}{8\rho} \int \frac{\de q}{(2\pi)^D}k_0^2(q)\left[\rho\Delta c(q)-1\right]\left[\left(\Delta c''(q)-\frac{\Delta c'(q)}{q}\right)\cos^2\theta+\frac{\Delta c'(q)}{q}\right]\\
-\frac{1}{8} \int \frac{\de q}{(2\pi)^D}k_0^2(q)\left(\Delta c'(q)\right)^2\cos^2\theta
\end{gather}
where $\theta$ is the angle between the D-dimensional vector $q$ and one of the coordinate axis. In $D=3$ we get
\begin{gather}
\sigma=\frac{1}{48\pi^2} \int_0^\infty \de q\, k_0^2(q)\left\{\frac{1}{\rho}\left[\rho\Delta c (q)-1\right]\left[q^2\Delta c''(q)+2q\Delta c'(q)\right]-q^2\left(\Delta c'(q)\right)^2\right\} \ .
\end{gather}
From Eq.~(\ref{midef}) we can compute also the parameters $m_2$ and $m_3$.
The expressions for these two quantities are
\beq\begin{split} \label{m2_m3_HNC}
m_2&=- \int \frac{\de q}{(2\pi)^D}k_0^2(q)\tilde c(q)\left[\frac{1}{\r}-\Delta c(q)\right]\\
m_3&=\frac{1}{2} \int \frac{\de q}{(2\pi)^D}k_0^2(q)\tilde c^2(q)\:.
\end{split}\eeq

\subsection{Summary of the numerical results}

Using the HNC approximation for the free-energy, i.e. neglecting 2PI diagrams
in Eq.~(\ref{GammaGen}), leads to a self-consistent equation for the order parameter
$\r^{(2)}_{ab}$ via Eq.~(\ref{extremum_g}). With the choice of a replica-symmetric structure
Eq.~(\ref{SP}), and setting $m=1$, we obtain a self-consistent equation for the diagonal correlations
that coincides with the liquid HNC equation, and a
self-consistent equation for the off-diagonal parts that takes the diagonal ones as input 
(see~\cite{PZ10} for details). Note that although this calculation is possible in any $D$, here
we restrict to $D=3$ for simplicity.

The equations are solved numerically using a standard Picard iteration 
scheme. We first focus on the diagonal part. We start from a very
low density $\rho \approx 0.2$ and gradually increase the density while
following the evolution of the solution.
At high enough density, if we solve the off-diagonal equation starting 
from a suitable guess for the off-diagonal $\tilde{c}$, we obtain a non-trivial solution.
When a non-trivial $\tilde{c}$ has been obtained for a given $\rho$, 
we gradually lower the density while following the evolution of
$\tilde{c}$, in order to get very close to the critical
point $\r_d$ where the solution disappears.

\begin{figure}[htb]
\includegraphics[width=.42\textwidth]{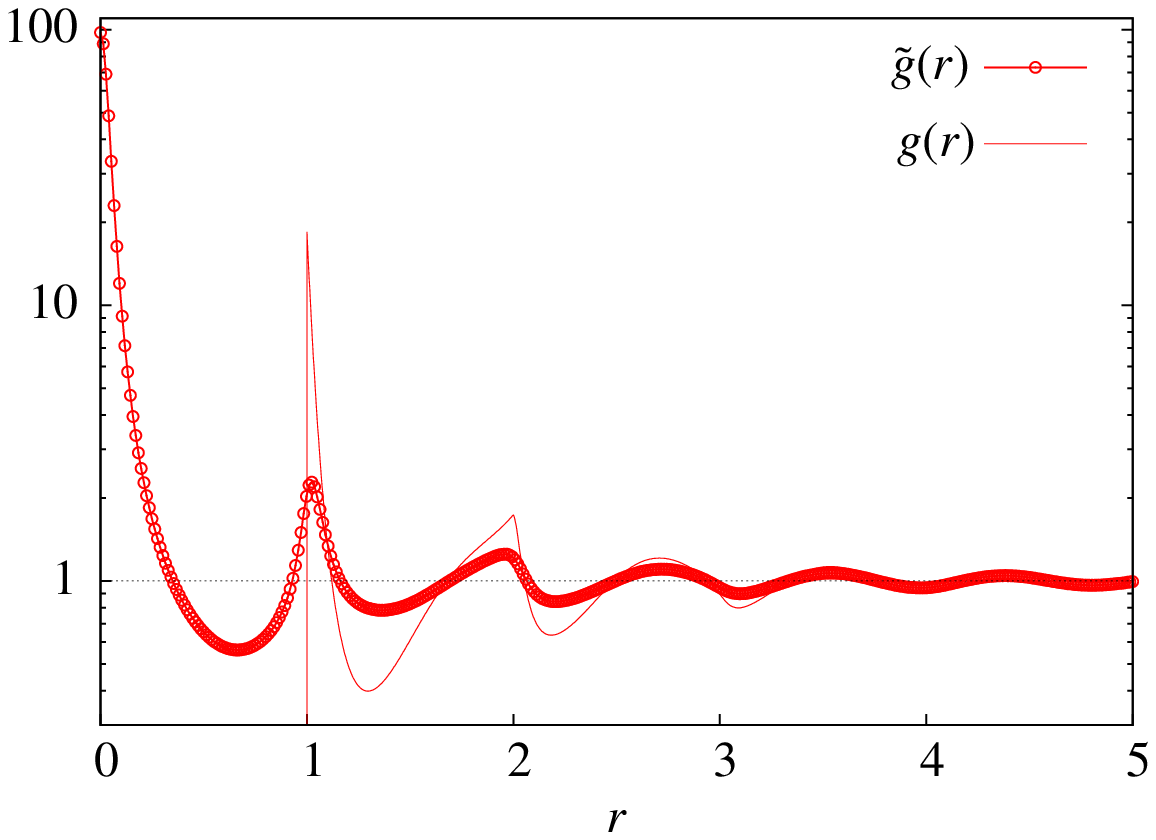}
\includegraphics[width=.45\textwidth]{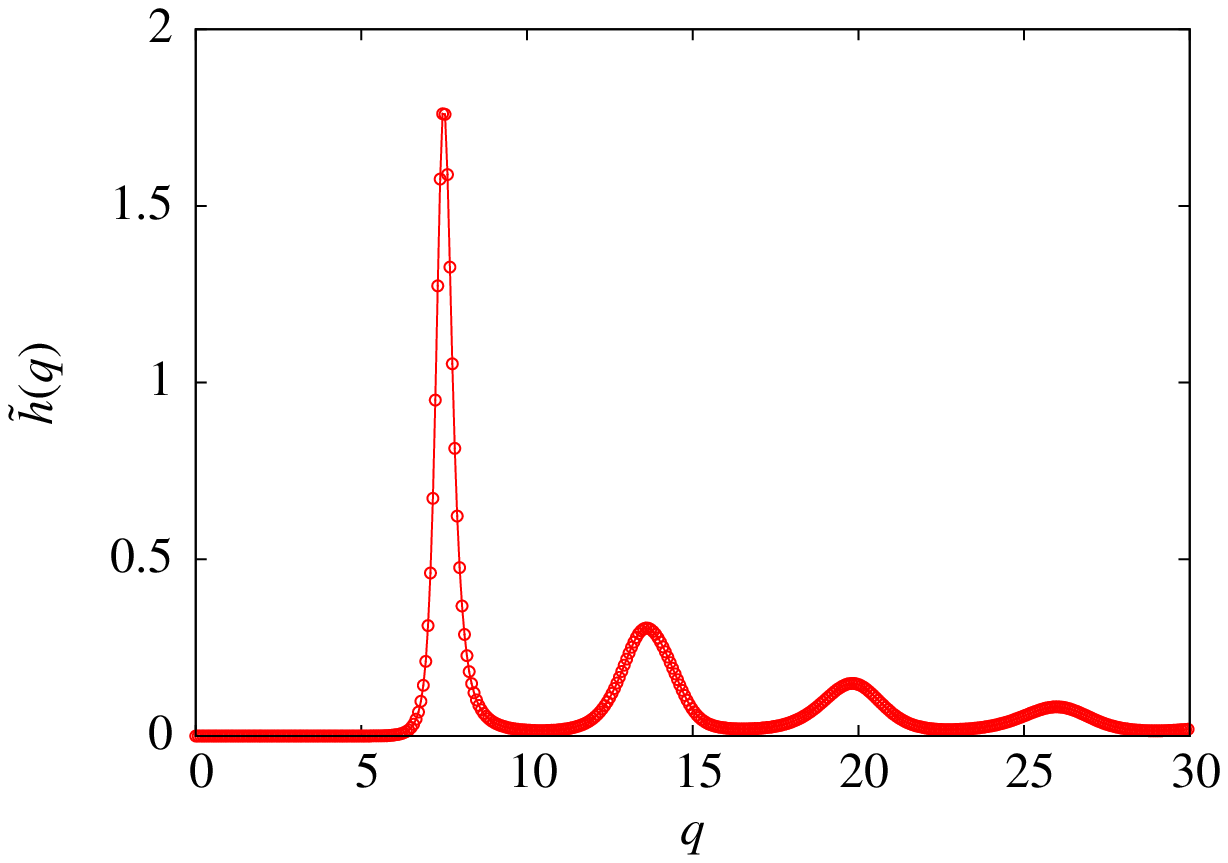}
\includegraphics[width=.45\textwidth]{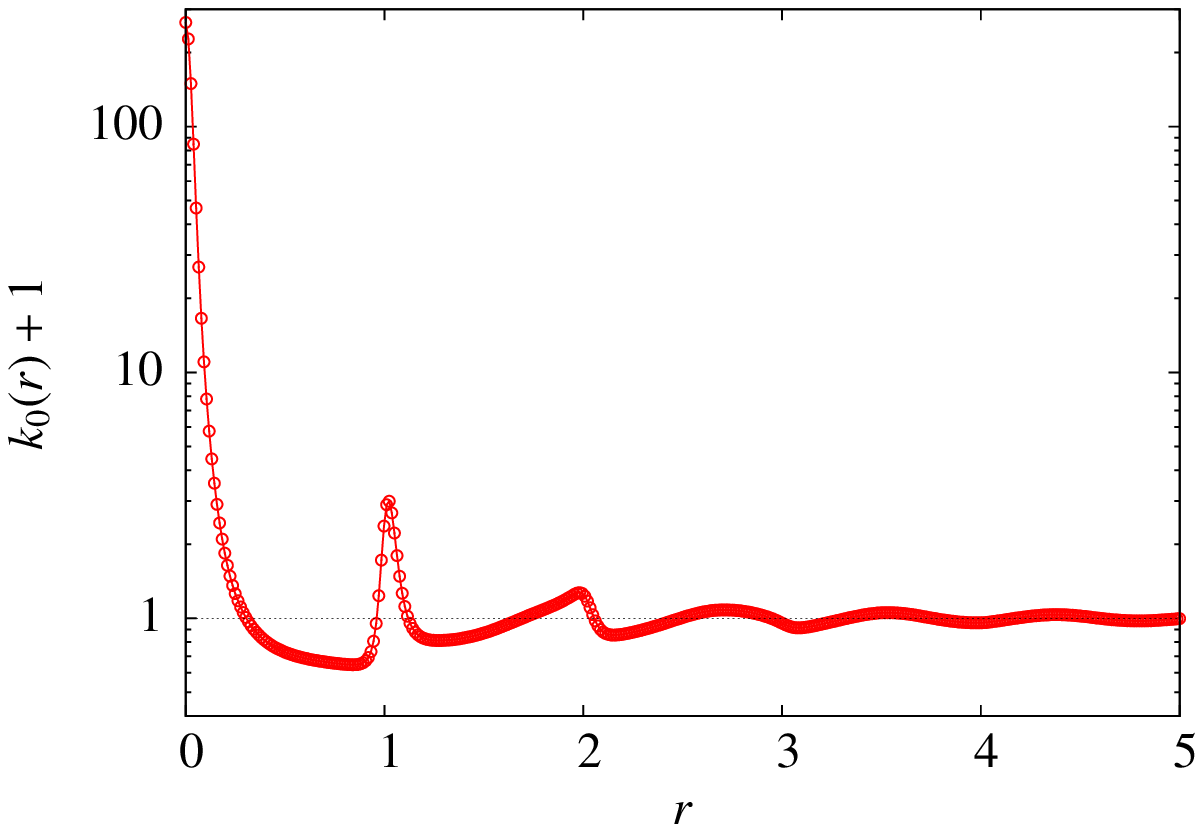}
\includegraphics[width=.45\textwidth]{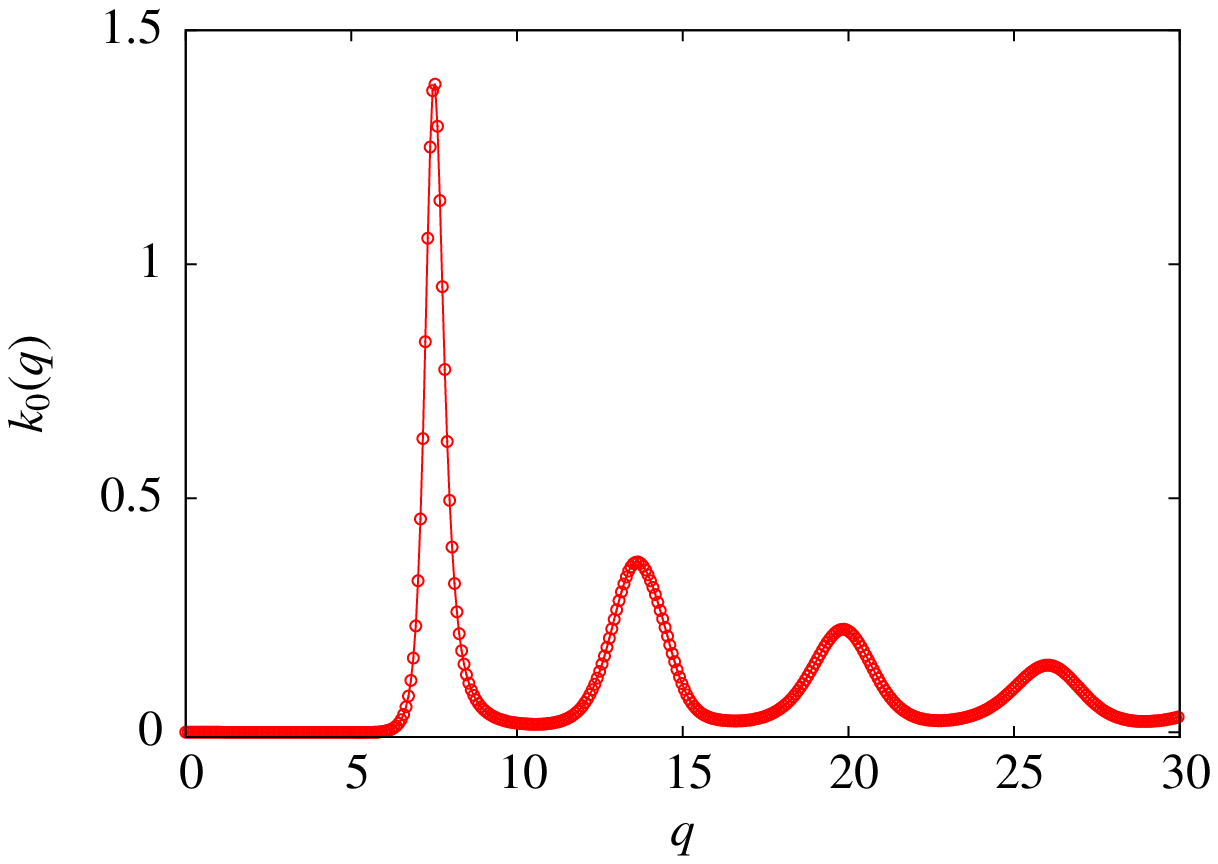}
\caption{
Off-diagonal correlations and the zero mode, all computed at $\r_d$ for hard spheres in $D=3$.
}
\label{fig:HNC_plots}
\end{figure}

For the obtained values of $\rho>\rho_d$, we then numerically compute the
derivative of the order 
parameter $\wt g$ with respect to density, both in Fourier and
real space. Recalling that this derivative is divergent at the transition, as
shown in Eq.~(\ref{sqrtSP}), we determine the precise
value of $\rho_d$ by imposing that for $k \approx 2 \p$ in Fourier space and
$r=0$ in real space, the
derivative of the correlation function scale as $\sqrt{\rho-\rho_d}$, using
$\rho_d$ as a fitting parameter. We used these particular values of $r$ and
$k$, since we observed 
that they were the most sensitive to density changes. The prefactor of this
square-root behavior is by definition the zero-mode $k_0$ in Eq.~(\ref{sqrtSP}). 
We check the validity of this scaling by computing $k_0$ at two different
densities slightly above $\rho_d$, and by checking that $k_0$ does not depend
on $\rho$ close enough to $\rho_d$. Note that here we included the constant $\k$
in $k_0$. In fact it is easy to see that the overall normalization of $k_0$ 
does not affect any of the physical observables.
We show in Figure \ref{fig:HNC_plots} the typical shape of the zero-mode $k_0$
in Fourier and real space, the off-diagonal pair correlation function
$\tilde{g}$ in real space, and its Fourier transform, all computed at $\r_d$.

We applied this numerical scheme to several benchmark models of monodisperse
systems of three-dimensional spherical particles that have pair-wise
interactions.
The models we used are:
\begin{itemize}
\item Hard spheres (HS):
\beq
v(r) = \left\{ \begin{array}{ll}
\io & \text{if } r < r_0 \\
0 & \text{if } r > r_0
\end{array} \right. ,
\eeq 
\item Harmonic spheres (HarmS):
\beq
v(r) = \left\{ \begin{array}{ll}
\e (1-r/r_0)^2 & \text{if } r < r_0 \\
0 & \text{if } r > r_0
\end{array} \right. ,
\eeq
for various temperatures (note that for $\beta \to \io$ this
potential reduces to the hard-spheres potential),
\item Lennard-Jones (LJ):
\beq
v(r) = 4 \e \left[ \left(\frac{r_0}{r}\right)^{12} - \left( \frac{r_0}{r} \right)^6 \right] ,
\eeq
\item Weeks-Chandler-Andersen (WCA):
\beq
v(r) = 4 \e \left[ \left(\frac{r_0}{r}\right)^{12} - \left( \frac{r_0}{r} \right)^6 +\frac14 \right] \th(r_0 2^{1/6} - r) ,
\eeq
\item Soft-spheres (SS):
\beq
v(r) =  \e \left(\frac{r_0}{r} \right)^n 
\eeq
for $n=6,9,12$.
\end{itemize}
The lengths are computed in units of the particle diameter $r_0$, and the temperatures in unit
of the energy scale of the potential $\e$ (except for HS, for which temperature is irrelevant).
Note that for SS temperature and density are not
independent variables, so we used the density as a control parameter.

Once $k_0$, $c$ and $\tilde{c}$ have been determined at the critical point, we
can readily compute the different parameters involved in the calculation of
the exponent parameter, the prefactor of the correlation length, the prefactor
of the Ginzburg criterium, as well as the prefactor of the divergent part of
the four-point correlation function. Note however that the latter depends on
the function $f$ that we choose in Eq.~(\ref{def_order_parameter}) to define our order parameter. In the numerical computation we used a box function
\beq
f(x)= (2A)^{-D} \prod_{i=1}^D \th(A^2 - x_i^2) \ ,
\eeq
where $\th(x)$ is the Heaviside step function. We choose $A=0.1 r_0$.
For all systems, we give the four main results of this paper in Tables~\ref{tab1} and \ref{tab2}:
\begin{itemize}
\item The value of $\l$ given by Eq.~(\ref{lambda_HNC}).
\item The prefactor $\xi_0 = \sqrt{\sigma/\mu}$ of the correlation function $\xi = \xi_0 \ee^{-1/4}$,
see Eq.~(\ref{xiGth}).
\item The prefactor of the divergent part of the four-point correlation function $G_{th}$ as given in Eq.~(\ref{4pointfunction_prefactor}).
\item The prefactor of the Ginzburg criterion given by Eqs.~\eqref{ginzburg} and \eqref{mass^2}.
\end{itemize}
Moreover let us recall that the quantities reported in the tables are not exactly the quantities defined in the previous sections. 
In particular the coefficients $\mu$ and $\sigma$ and the two masses $m_2$ and $m_3$ are computed by including in $k_0$ the normalization factor $\kappa$
(hence they are multiplied by $\kappa^2$)
and moreover the coefficients $w_1$ and $w_2$ are the ones given by (\ref{w1_w2}) but again including $\kappa$ in $k_0$ (hence they are multiplied
by $\kappa^3$) and
without the $f \star k_0$ factor which clearly does not enter in the computation of $\l$. In any case the values of the four physical quantities
$\l, \xi_0, G_0,$ Gi, are independent of these normalizations.
We found that, for all these systems, the dependance of the results on the real and Fourier
space cut-offs that are needed for the numerical calculation is significant, 
altering the results on our predictions to
within $10^{-3}$. 
For the case of hard-spheres, we checked that the results
become increasingly stable when diminishing both cut-offs simultaneously. The
error that we make because of the finite cutoffs are however not very important
because numerical simulations and experiments can not, for the moment, provide
results with a better accuracy, due to the difficulty in
accessing the critical region close to the glass transition.

Our results for $\l \approx 0.35$ are quite different from the ones obtained from Mode-Coupling Theory~\cite{Be86,BGL89,Go09},
which finds $\l \approx 0.7$ in very good agreement with numerical simulations. 
This confirms earlier indications, that the replicated HNC approximation
is not a good scheme for quantitative calculations~\cite{PZ10}. 
Note that the values reported in a preliminary report on this work~\cite{FJPUZ12}
were missing a factor of $1/2$ in the expression of $w_2$ in Eq.~\eqref{w1_w2_HNC}.
This missing factor was found thanks to
an independent calculation of $\l$ from a completely different method~\cite{FPU13}. 
The factor affects both the value of $\l$ and
that of the Ginzburg number, and unfortunately, the correct value of $\l \approx 0.35$ from the replicated HNC approximation
turns out to be quite different from the one
obtained in numerical simulations $\l \approx 0.7$. However, the latter was accidentally coincident with the
value reported in~\cite{FJPUZ12} because of the missing factor of $1/2$, which of course was not helpful in finding the error.
A calculation of $\l$ in the small cage expansion
gives a much better agreement with the Mode-Coupling result~\cite{KPUZ13}.
Concerning the other observables,
unfortunately, not many numerical data for the behavior of the thermal correlation in the $\b$ regime are not available.
We have checked that our results are roughly consistent with the results of~\cite{SA08}, but more precise simulations would be very
useful to test our predictions.

\begin{table*}
\label{tab1}
\begin{tabular}{|cc|ccccccc|cccc|}
\hline
System & $T$ & $\rho_{\rm d}$ & $-w_1$ & $-w_2$ & $m_2$ & $m_3$ & $\sigma$ & $\mu$ & $\lambda$ & $\xi_0$ & $G_0$ & Gi   \\
\hline
SS-6  & 1 & 6.691 & 3.88$\cdot10^{-6}$ & 1.35$\cdot10^{-6}$  & -0.000925 & 0.000110 & 0.000195 & 0.000525 & 0.348 & 0.601  & 224  & 0.0267 \\
SS-9  & 1 & 2.912 & 0.0000772  & 0.0000272    & -0.00539  & 0.000633  & 0.00163  & 0.00543  & 0.353 & 0.548  & 34.3 & 0.0125\\
SS-12 & 1 & 2.057 & 0.000275  & 0.0000973    & -0.0116  & 0.00132  & 0.00378  & 0.0152  & 0.354 & 0.498  & 14.2 & 0.0118\\
LJ  & 0.7  & 1.407 & 0.00106  & 0.000376    & -0.0258  & 0.00290  & 0.00989   & 0.0414  & 0.355 &0.489   & 6.00 & 0.00833 \\
HarmS & $10^{-3}$ & 1.336  &   0.00129 & 0.000465 &    -0.0336 & 0.00343 & 0.00772 & 0.0779  & 0.359  & 0.315 & 2.82 & 0.0434\\
HarmS & $10^{-4}$ &  1.196   &  0.00165 & 0.000622 &  -0.0403 & 0.00386 & 0.00819 &  0.109 & 0.378   & 0.274 & 1.69 & 0.0632\\
HarmS & $10^{-5}$  & 1.170  &   0.00174 & 0.000663 &  -0.0416 & 0.00395 & 0.00845  & 0.109 & 0.382   & 0.278 & 1.66 & 0.0635\\
HS &  0   &  1.169  &   0.00174 & 0.000664 &  -0.0418 & 0.00397 & 0.00847 &  0.108 & 0.381 & 0.280 & 1.67 & 0.0639\\
\hline
\end{tabular}
\caption{
Numerical values of the coefficients of the effective action and the physical quantities from the HNC approximation. 
For each potential, lengths are given in units of $r_0$ and energies in units
of $\e$, with $k_B=1$. 
Data at fixed temperature, using density as a control parameter with $\ee = \r_{\rm d} - \r$.
}
\end{table*}

\begin{table*}
\label{tab2}
\begin{tabular}{|cc|ccccccc|cccc|}
\hline
System & $\r$ & $T_{\rm d}$ & $-w_1$ & $-w_2$ & $m_2$ & $m_3$ & $\sigma$ & $\mu$ & $\lambda$ & $\xi_0$ & $G_0$ & Gi   \\
\hline
LJ  & 1.2 & 0.336 & 0.00186 & 0.000663 & -0.0361 & 0.00403 & 0.0147 & 0.0572 & 0.356 & 0.507 & 4.56 & 0.00730\\
LJ & 1.27 & 0.438 & 0.00153 & 0.000541 & -0.0321 & 0.00370 & 0.0128 & 0.0447 & 0.353 & 0.536 & 5.74 & 0.00771\\
LJ  & 1.4 & 0.684 & 0.00108 & 0.000383 & -0.0260 & 0.00293 & 0.0100 & 0.0292 & 0.355 & 0.586 & 8.52 & 0.00825\\
WCA & 1.2 & 0.325 & 0.00195 & 0.000686 & -0.0389 & 0.00426 & 0.0133 & 0.0607 & 0.351 & 0.467 & 4.37 & 0.0134\\
WCA  & 1.4  & 0.692 & 0.00111 & 0.000388 & -0.0270 & 0.00301 & 0.00966 & 0.0291 & 0.350 & 0.576 & 8.67 & 0.0106\\
\hline
\end{tabular}
\caption{
Numerical values of the coefficients of the effective action and the physical quantities from the HNC approximation. 
For each potential, lengths are given in units of $r_0$ and energies in units
of $\e$, with $k_B=1$. 
Data at fixed density, using temperature as a control parameter with $\ee = T_{\rm d} - T$.
}
\end{table*}

\section{Conclusions}

A complete characterization of dynamical heterogeneities in the $\beta$ regime has been obtained using a static equilibrium approach in the framework of the replicated liquid theory. The criticality of the four point density correlation functions has been analyzed through the computation of the stability operator of the replicated Gibbs free energy around the glassy solution. This kernel operator has a soft mode that is responsible for the growth of the various type of susceptibilities at the dynamical transition. Having identified the soft mode, we have produced a gradient expansion for the field theory which describes the fluctuations of the two point density field that are along the zero mode itself. In this way we can focus on the critical part of the quantities we are interested in. Then we have studied the theory at the 
Gaussian level and we have performed a one loop analysis in order to see where the mean field regime breaks. This results in a Ginzburg criterion for the dynamical transition. Our approach is completely general and it relies only on the fact that the glassy phenomenology can be captured by the one-step replica symmetry breaking scheme. Moreover, to produce some quantitative predictions we have computed in the HNC approximation all the observables we are interested in: the correlation length, the thermal four points correlation function, the Mode-Coupling exponent parameter and all the couplings of the effective replica field theory. 

Our calculations can be straightforwardly extended to other more accurate approximation schemes and this is one of the points that must be explored. The question about how to attack systematically the problem in the $\alpha$ regime remains open. One might think that replicas are useful mostly in the $\beta$ regime where one is exploring the interior of one metastable state, 
while the study of the barrier-crossing $\alpha$ regime requires a full dynamical approach. However recently it has been discussed how replicas can be used to obtain the full long time reparametrization invariant dynamics~\cite{FP12}. This is surprising at first sight because it gives a recipe to obtain some results for the dynamics directly from the statics; however the time sector that it explores is the reparametrization invariant one where quasi-equilibrium holds. It would be interesting to see what is the insight that can be gained for dynamical heterogeneities in the $\alpha$ regime from the application of this line of reasoning to the replicated liquid case.

Finally, a recent numerical investigation suggested the presence of an upper critical dimension $d_u=8$ for glassy dynamics~\cite{CPZ12}.
A more compete comparison between the theoretical predictions obtained here and the numerical results would be very helpful to understand the nature of the corrections to mean
field that become important below~$d=8$.

\acknowledgments
We thank G.~Biroli, J.-P.~Bouchaud, P.~Charbonneau, A.~Ikeda, and D.~R.~Reichman for very useful discussions.
H. Jacquin PhD work is funded by a Fondation CFM-JP Aguilar grant.
GP and PU acknowledge financial support from
the European Research Council through ERC grant agreement
No.~247328.

\appendix

\section{Double Counting Problem on the $\f^4$ theory} 
\label{app:DC}

Let us consider the $\f^4$ theory defined by the action in Eq.~\eqref{eq:Sbare1}.
The generating functional is defined by Eq.~\eqref{eq:GammaLegendreS}.
At the mean field level we have $\G_{\rm MF}[\phi] = S[\phi]$. Including
one loop corrections, we obtain the following expression:
\beq
\Gamma_{\rm 1L}[\phi]= \G_{\rm MF}[\phi] + \frac 12 \mathrm{Tr}\log K \ ,
\eeq
where $K$ is the following operator
\beq
K(x,y)=\delta(x-y)(m_0^2-\nabla^2+\frac{g}{2}\phi^2(x)) \ .
\eeq
By isolating a free operator $K_0(x,y)=\delta(x-y)(m_0^2-\nabla^2)$ 
and defining an operator $(\phi^2)(x,y) = \phi^2(x)\delta(x-y)$
we can rewrite the $K$ operator in the following way
\beq
K=K_0\otimes[1+\frac{g}{2}K_0^{-1}\otimes \phi^2] \ ,
\eeq
where $\otimes$ is the operator (integral) product. Moreover
\beq
K_0^{-1}(x,y)=\int \frac{\de p}{(2\pi)^D}\frac{e^{ip(x-y)}}{p^2+m_0^2}
\eeq
It follows that
\beq
\Gamma_{\rm 1L}[\phi]=\G_{\rm MF}[\phi] + \frac 12 \mathrm{Tr}\log \left[1+\frac{g}{2}K_0^{-1}\otimes \phi^2 \right]+
\frac 12 \mathrm{Tr}\log K_0 \ .
\eeq
The last term is $\phi$-independent which means that it can be neglected for our purposes. 
In fact we want to see what happens if we compute the two point function at order $g$ using 
$\Gamma_{\rm 1L}$ as an action instead of using the bare action $S$. 
Because we want the two point function at order $g$ we need to perform a 
Taylor expansion of the extra term $\mathrm{Tr}\log [1+(g/2)K_0^{-1}\otimes \phi^2]$ which is given by
\beq\label{eqapp:1Lcorr}
\mathrm{Tr}\log \left[1+\frac{g}{2}K_0^{-1}\otimes \phi^2\right]= \mathrm{Tr} \left[ \frac{g}{2} K_0^{-1}\otimes \phi^2 \right]+O(g^2)=
\frac{g}{2}D_1(m_0^2)\int \de x \phi^2(x)+O(g^2)
\eeq
where
\beq
D_1(m_0^2)=\int \frac{\de p}{(2\pi)^D}\frac{1}{p^2+m_0^2}
\eeq
We now use $\G_{\rm 1L}$ as the bare action, $S_{\rm 1L}[\f] = \G_{\rm 1L}[\f]$ and
compute the two point function. The diagrammatic rules are the standard ones for the $\f^4$ theory~\cite{ParisiBook}:
\beq\label{GDCdiag}
G(p) =
\begin{picture}(30,15)(-30,-2)
\SetColor{Black}
\SetWidth{1}
\Line(-15,0)(15,0)
\end{picture}
\ \ \ \ \ \ \ \ +
\begin{picture}(30,15)(-30,-2)
\SetColor{Black}
\SetWidth{1}
\Line(-15,0)(15,0)
\CCirc(0,0){2}{Black}{White}
\end{picture}
\ \ \ \ \ \ \ \ +
\begin{picture}(30,30)(-30,-2)
\SetColor{Black}
\SetWidth{1}
\Line(-15,0)(15,0)
\CArc(0,6)(6,0,360)
\end{picture}
\eeq
where the second diagram is the one originated by the term in Eq.~\eqref{eqapp:1Lcorr}. 
Note that the second diagram has exactly the same expression as the last one. This is the expression of the double counting problem.

At this point we can compute the Ginzburg Criterion starting by $S_{\rm 1L}[\f] = \G_{\rm 1L}[\f]$.
The action at order $g$ is then given by
\beq
S_{\rm 1L}[\f]
 = \frac 12\int \de x \f(x)\left[ -\nabla^2+ m_0^2 + \frac{g}{2}D_1(m_0^2)  \right]\f(x)+
\frac{g}{4!}\int \de x \f^4(x) \ .
\eeq
It should be clear at this point that the only difference with the bare action $S[\f]$ is a change
of the bare mass, $m_0 \to  m_0^2 + \frac{g}{2}D_1(m_0^2)$. However, we have seen in Section~\ref{sec:illD}
that the final expression of the Ginzburg criterion is expressed in terms of the renormalized mass only,
and is therefore not affected by a change of the bare mass. We conclude that whatever microscopic action we use
-- provided it can be developed in powers of $\f^2$ at small $\f$, which is the crucial assumption of mean field theory --
will give the same results for the Ginzburg criterion.
In this sense the Ginzburg criterion can be thought as a check {\it a posteriori} of this assumption.

\bibliographystyle{mioaps}
\bibliography{HNCfield}

\end{document}